# LOCAL BIFURCATION ANALYSIS OF CIRCULAR VON-KÁRMÁN PLATE WITH KIRCHHOFF ROD BOUNDARY*

DEEPANKAR DAS† AND BASANT LAL SHARMA‡

**Abstract.** Symmetry based reduction is applied to the buckling of a circular von-Kármán plate with Kirchhoff rod boundary, where a mismatch between the edge length and the perimeter of plate is treated as the bifurcation parameter. A nonlinear operator formulation describes the equilibrium of the elastic rod-plate system. The critical points, as potential bifurcation points, are stated using the linearized operator. The symmetry of null space for each critical point is identified as a subgroup of the complete symmetry group of nonlinear problem; the equivariance associated with the nonlinear operator is used in this process. Sufficient evidence is provided for each critical point to be a bifurcation point for the symmetry-reduced problem and post-buckling analysis is carried out using Lyapunov–Schmidt reduction. Bifurcation curves are obtained up to quadratic order in bifurcation parameter away from each critical value. Theoretical results for bifurcation curves are validated against the numerical simulation based on a symmetry-reduced finite element method for some illustrative examples of critical points. A numerical study is carried out for the dependence of the coefficient of quadratic term in the bifurcation parameter when structural parameters are varied in a neighbourhood of four fixed sets of structural parameters. Numerical results based on a symmetry-reduced finite element analysis confirm that the nonlinear solution agrees with the local theoretical behaviour close to a critical point but deviates further away from it. Using these tools, two main conclusions are reached. First it is observed that the critical points of the linearized problem are indeed bifurcation points. Second, an alteration in the nature of bifurcation is observed during the parameter sweep study when the plate is in tension. That is, near certain values of the structural parameters, a supercritical pitchfork bifurcation locally evolves into a subcritical pitchfork.

**Key words.** von-Kármán plate, Kirchhoff rod, post-buckling, local bifurcation, first variation

**MSC codes.** 74B20, 74G35, 37G40.

**Introduction.** The study of bifurcation and stability in slender elastic structures has been extensively addressed in the literature [6, 17, 25, 18, 3, 16]. Evidently, a nonlinear system of partial differential equations with symmetries often exhibits bifurcation. The foundational theories in local analysis, for this purpose, particularly, the symmetry-based methods [24] and the relevant Lyapunov-Schmidt reduction [27], provide the groundwork for present study. Several research findings such as those reported in [14], [10], and [26] demonstrate the role of symmetry in analyzing bifurcation behaviors in various mechanical contexts. Applications of these theoretical tools to closely related problems in solid mechanics are discussed in [11, 12, 13, 32, 34, 31, 23], for example. Being cognizant of a range of analyses available in the existing literature, which are closely related to specific model system considered in our study, there are several other existing studies which have explored stability and bifurcation in systems involving structures of different dimensionalities. For example, the question of stability and bifurcations in a soap film spanning an elastic loop is investigated in [4], while [1] examined a growing rod in two dimensions and identified transitions between supercritical and subcritical bifurcation behaviors. The latter observations are indeed similar to those found in this article. The interaction between a rod and a confining circular tube is explored in [19], drawing parallels to the examination of structural interactions in current research work. Additionally, models of extensible rods, such as those used in [21], could serve as a basis for extending the current work by addressing more complex buckling and post-buckling scenarios.

†Department of Mechanical Engineering, Indian Institute of Technology Kanpur, Kanpur, U.P., 208016, India (deepankd@iitk.ac.in, http://www.home.iitk.ac.in/~deepankd/).

‡Department of Mechanical Engineering, Indian Institute of Technology Kanpur, Kanpur, U.P., 208016, India (bls@iitk.ac.in, http://home.iitk.ac.in/~bls/Homepage/Home.html/)

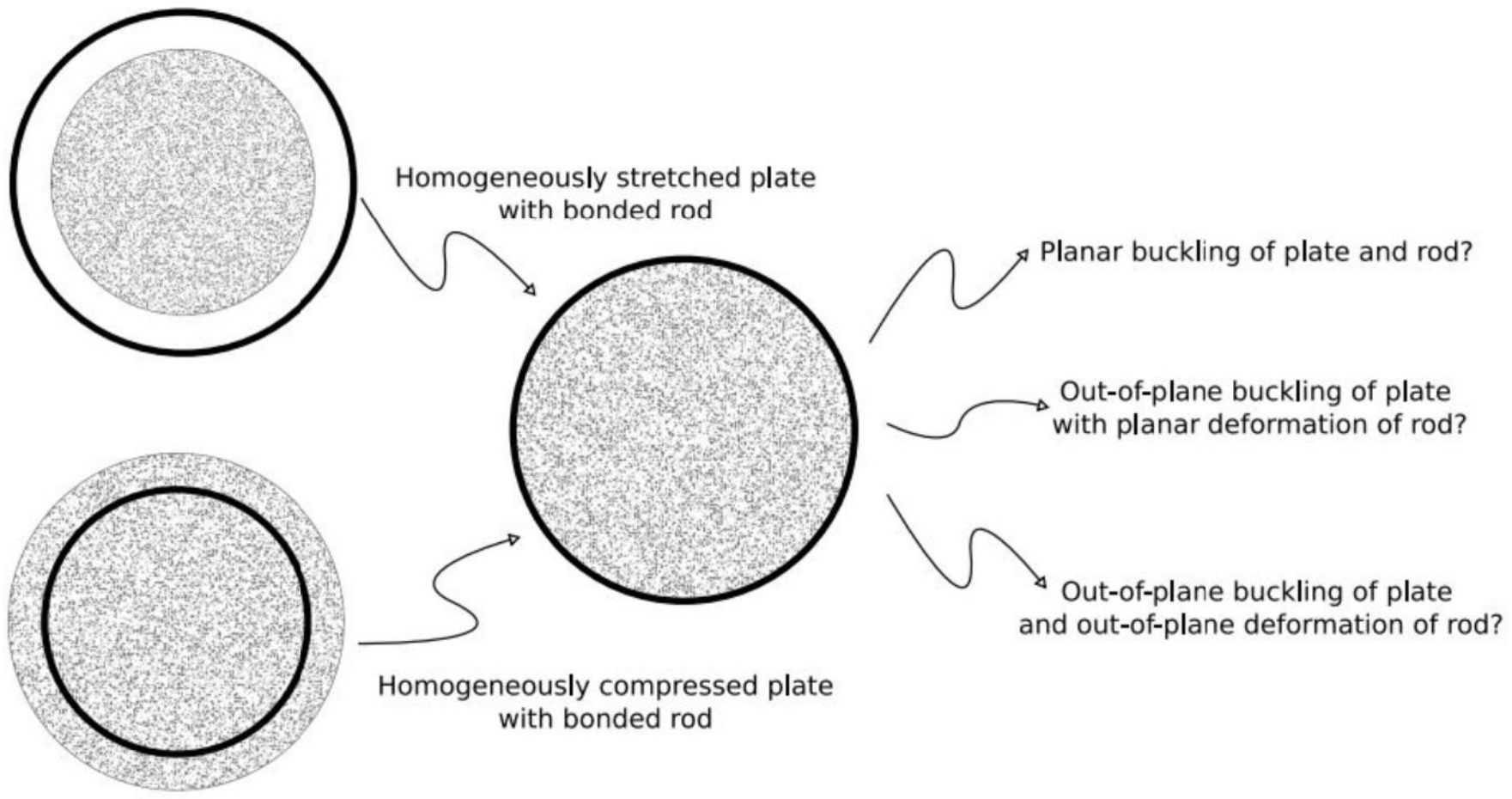

Fig. 1: A schematic drawing illustrating the bonding between rod (in black shade) and the boundary of plate (in granular gray shade) leading to a self-stressed rod-plate system which can lead to a variety of buckling instabilities.

In this article, the symmetry-based reduction is applied to the buckling of a circular (von-Kármán) elastic plate with elastic (inextensible and unshearable special Cosserat) rod boundary. The mismatch between the length of rod, in its stress-free configuration, and the perimeter of the plate, also in its stress-free configuration, is represented as the bifurcation parameter. See Fig. 1 for a schematic of such rod-plate system. The structural interaction between the plate and the rod is assumed to be, for simplicity, such that the displacement is continuous across the joint and the edge-moment is zero. In other words, the plate is joined to the rod in such a way that only the distributed forces (Kirchhoff shear, membrane forces) are transferred, while the (edge-)bending moment vanishes and the (edge-)torsional moment is balanced solely through force interactions; overall, the moments in plate at the boundary are not directly related to the internal moments in the rod. From a statical perspective, this means that the equilibrium between the plate and rod is maintained through force interactions alone, making it a moment-free connection; see details in [5]. Expanding on the nature of joint between the plate and the rod [5], it is useful to note that the out-of-plane displacement of the plate boundary coincides with that of the deformed rod center-line, and the edge-moment is zero, providing two scalar boundary conditions for the fourth order partial differential equation, effectively, for determination of out-of-plane displacement, while the in-plane displacement of the plate boundary coincides with that of the rod center-line providing two scalar boundary conditions for the two scalar second order partial differential equations, effectively, for the two components of in-plane displacement; in general, the three scalar equations are coupled. The trivial solution corresponds to a homogeneously deformed plate (as shown in Fig. 1) in presence of the inextensible circular rod boundary bonded in aforementioned manner, as the base solution, and the question of finding solution branches, bifurcating from this solution, as the bifurcation parameter varies, is of central interest. Indeed, as a result of the nature of joint between the inextensible circular rod and the edge of the extensible plate, the rod-plate system develops internal stress (self-stress) in the natural planar configuration, so that for some critical values of the mismatched perimeters, as expected, buckled states arise which are either dominated by the plate deformation or dominated by the rod deformation. In [5], we have formulated the rod-plate equilibrium problem as a nonlinear operator and conducted the critical point analysis to identify all candidate values (of mismatch parameter) for bifurcation and their associated symmetries. In this work, we analyze the existence of bifurcation and use the results of [5] to obtain the post buckling behaviour of the nonlinear problem by using those symmetries. This problem is qualitatively motivated by applications in engineering and sciences, for example, composite structures, biological systems, etc. Potential applications of such analyses are anticipated in biological and soft robotic systems. For instance, models with the twisting growth of roots are studied by [28], while rod buckling driven by internal fields to model flagellum-like structures is investigates by [20]. The issue of tunable buckling strength of elastic shells is discussed in [35] while the use of buckling for actuation in soft machines is presented in [36]. Indeed, these are only few examples out of many available in the literature, which highlight the relevance of the presented analysis in emerging fields of science and engineering.

**Outline.** After fixing notation, in §1 we provide the mathematical formulation of bifurcations in a plate bonded with rod. The linearization and critical points for in-plane as well as out-of-plane perturbed solution relative to the base solution are summarized in §2. In §3, Lyapunov–Schmidt reduction is applied; in particular, symmetry groups for equivariance are identified and described in §3.1. The symmetry-reduced Lyapunov–Schmidt technique is used in §4 to guarantee that criticial points are indeed bifurcation points; particularly in §3.3, symmetry-reduced problem is formulated, while sufficient evidence for bifurcation is provided in §4.1. The numerical results are presented in §5 where theoretical results are validated against numerical results based on symmetry-reduced finite element scheme; the scheme is summarized in Appendix §D. Other three appendices in the article before the appearance of references deal with the adjoint operator of the linearized operator and the relevant null spaces.

**Notation and mathematical preliminaries.** The set of real numbers (resp. positive reals) is denoted by $\mathbb{R}$ (resp. $\mathbb{R}^+$). We also use the symbol $\mathbb{R}_{2\pi}$ to represent periodic interval of length $2\pi$. Let $O$ denote the origin in three dimensional physical space (Euclidean point space) $\mathtt{E}$. Relative to $O$, the physical space $\mathtt{E}$ has structure of three dimensional Euclidean (vector) space which is denoted by $\mathbb{R}^3$. The standard inner-product is denoted by $\boldsymbol{a}\cdot\boldsymbol{b}$ for given $\boldsymbol{a},\boldsymbol{b}\in\mathbb{R}^3$; the cross-product is denoted by $\boldsymbol{a}\wedge\boldsymbol{b}$ for given $\boldsymbol{a},\boldsymbol{b}\in\mathbb{R}^3$. Let $\{\boldsymbol{e}_1,\boldsymbol{e}_2,\boldsymbol{e}_3\}$ be a fixed (right handed) orthonormal basis, i.e. standard basis, of $\mathbb{R}^3$; $\boldsymbol{a}=a_i\boldsymbol{e}_i$, $\boldsymbol{a}\in\mathbb{R}^3$, using indicial summation convention. The second order tensor $\boldsymbol{a}\otimes\boldsymbol{b}$, for given $\boldsymbol{a},\boldsymbol{b}\in\mathbb{R}^3$, is defined as a linear transformation from $\mathbb{R}^3$ to $\mathbb{R}^3$ such that $(\boldsymbol{a}\otimes\boldsymbol{b})[\boldsymbol{c}]=\boldsymbol{a}(\boldsymbol{b}\cdot\boldsymbol{c})$ for all $\boldsymbol{c}\in\mathbb{R}^3$; $\mathtt{I}:=\boldsymbol{e}_i\otimes\boldsymbol{e}_i$ represents the identity tensor. The transpose of a second order tensor $\mathbf{A}$ is denoted by $\mathbf{A}^\top$ and is defined by $\mathbf{A}\boldsymbol{u}\cdot\boldsymbol{v}=\mathbf{A}^\top\boldsymbol{v}\cdot\boldsymbol{u}$, for all $\boldsymbol{u},\boldsymbol{v}\in\mathbb{R}^3$; also $(\boldsymbol{a}\otimes\boldsymbol{b})^\top=\boldsymbol{b}\otimes\boldsymbol{a}$. The notation $\mathrm{skw}(\boldsymbol{a})$ stands for the skew tensor with axial vector $\boldsymbol{a}$, that is $\mathrm{skw}(\boldsymbol{a})[\boldsymbol{b}]=\boldsymbol{a}\wedge\boldsymbol{b}, \forall\boldsymbol{a},\boldsymbol{b}\in\mathbb{R}^3$. $\mathbf{A}=A_{ij}\boldsymbol{e}_i\otimes\boldsymbol{e}_j$, for a second order tensor $\mathbf{A}$. Inner product for second order tensors is defined as $\mathbf{A}:\mathbf{B}=A_{ij}B_{ij}$, for arbitrary second order tensors $\mathbf{A}$ and $\mathbf{B}$; also, $\boldsymbol{a}\otimes\boldsymbol{b}:\boldsymbol{c}\otimes\boldsymbol{d}=(\boldsymbol{a}\cdot\boldsymbol{c})(\boldsymbol{b}\cdot\boldsymbol{d})$. The second order tensors are typically referred as tensors. We follow the standard notation in continuum mechanics and solid mechanics [8]. Some non-standard notational choices appear as and when required. The square brackets, in general, denote linear action of an operator (say a tensor, first variation of functional, etc.) as clear from the context with an exception of use of similar notation for a closed interval of real line; for example, $\mathbf{A}\boldsymbol{a} = \mathbf{A}[\boldsymbol{a}] = A_{ij}a_j\boldsymbol{e}_i$. Given $\mathtt{R} \in \mathrm{SO}(3)$, the group of all rotation tensors in three dimensions, $\mathtt{R}\boldsymbol{a} \cdot \mathtt{R}\boldsymbol{b} = \boldsymbol{a} \cdot \boldsymbol{b}$ and $\mathtt{R}\boldsymbol{a} \wedge \mathtt{R}\boldsymbol{b} = \mathtt{R}(\boldsymbol{a} \wedge \boldsymbol{b})$ for all $\boldsymbol{a}, \boldsymbol{b} \in \mathbb{R}^3$. Rotation tensors are parametrized by the Gibbs vector [29]. Several identities of vector and integral calculus are used in various parts of this article. Besides vectors and tensors in three dimensions, we also employ analogous definitions for their restriction to $\boldsymbol{e}_1$-$\boldsymbol{e}_2$ plane. The two dimensional Euclidean space as $\mathrm{span}(\boldsymbol{e}_1, \boldsymbol{e}_2)$ is denoted by $\mathbb{R}^2$, considered as a part of $\mathbb{R}^3$. The inner-product on $\mathbb{R}^2$ is denoted by the same dot as in $\mathbb{R}^3$ and is defined as the restriction of dot product on $\mathbb{R}^3$. Symbol $\mathbf{I} := \boldsymbol{e}_1 \otimes \boldsymbol{e}_1 + \boldsymbol{e}_2 \otimes \boldsymbol{e}_2$ is the identity tensor on the plane $\mathbb{R}^2$. A tensor $\mathbf{A}$ on $\mathbb{R}^2$ is a symmetric tensor provided that $\mathbf{A}\boldsymbol{a} \cdot \boldsymbol{b} = \mathbf{A}\boldsymbol{b} \cdot \boldsymbol{a}$ for all $\boldsymbol{a}, \boldsymbol{b} \in \mathbb{R}^2$; the set of symmetric tensors on $\mathbb{R}^2$ is denoted by Sym. $\mathcal{C}^k(A, \mathbb{R})$ denotes the $k$ times continuously differentiable functions from $A \to \mathbb{R}$ for a compact set $A$ in $n$ dimensional Euclidean space.

**1. Problem definition.** We consider an inextensible and unshearable special Cosserat (Kirchhoff) rod, in the shape of a circular loop of diameter $\overline{\mathfrak{D}}$, with bending modulus $\overline{\beta}$ and twisting modulus $\overline{\gamma}$ bonded to a plate with stress-free diameter $\lambda\overline{\mathfrak{D}}$, thickness $\overline{h}$, Young's modulus $\overline{E}$, Poisson's ratio $\nu$, as briefly described in introduction (see Fig. 1). For the physical rod-plate system, the dimensionless *structure parameters* are $\nu, h, \mathfrak{D}, \beta, \gamma$, out of which the latter four are defined such that

$$h := \frac{\overline{h}}{L}, \quad \mathfrak{D} := \frac{\overline{\mathfrak{D}}}{L}, \quad \beta := \frac{\overline{\beta}}{\mathfrak{C}L^3}, \quad \gamma := \frac{\overline{\gamma}}{\mathfrak{C}L^3}, \text{ where } \mathfrak{C} := \frac{\overline{E}\,\overline{h}}{12(1-\nu^2)} \tag{1.1}$$

where $\overline{\beta}$ and $\overline{\gamma}$ are the bending modulus and torsional modulus of the rod, respectively, and $L$ is the length scale that can be chosen in multiple ways. The Kirchhoff rod is assumed to be isotropic and homogeneous, so that $\overline{\beta}$ and $\overline{\gamma}$ depend on the cross-section and material properties. In this article, we choose $L$ such that $L := \frac{1}{2}\overline{\mathfrak{D}}$ (as an example of another choice, we mention $L$ can be also chosen as $(\overline{\beta}/\mathfrak{C})^{1/3}$), so that the scaled rod center-line in stress-free configuration of rod is assumed to be a unit circle. However, for clarity and to keep connection with the physical problem, we do not remove $\mathfrak{D}$ from the ensuing mathematical expressions. The physical problem of rod-plate system is discussed with all details in [5].

Let $\Omega$ (resp. $\overline{\Omega}$) be an open (resp. closed) circular disc of diameter $\mathfrak{D}$ centered at origin. This circular disc represents the homogeneously deformed plate (as shown in schematic Fig. 1). We use $\Gamma$ in place of $\partial\Omega$ to represent boundary of $\Omega$. To describe the formulation of the mechanics of rod-plate problem, we consider the following space of functions on $\overline{\Omega}$:

$$\mathcal{U} := (\mathcal{C}^1)^2 \times \mathcal{C}^2 \times (\mathcal{C}^1)^3 \times (\mathcal{C}^2)^3 \times (\mathcal{C}^1)^2 \times (\mathcal{P}^2)^3 \times (\mathcal{P}^1)^3, \tag{1.2}$$

where $(X)^n$ represents the Cartesian product $X \times \ldots \times X$ ($n$ times) of some space $X$,

$$\mathcal{C}^k := \mathcal{C}^{k,\alpha}(\overline{\Omega}, \mathbb{R}) \text{ and } \mathcal{P}^k := \mathcal{C}^{k,\alpha}(\Gamma, \mathbb{R}), \quad k = 0, 1, 2, \ldots, \tag{1.3}$$

(with $\alpha \in (0, 1]$ as the Hölder exponent [22]) and a typical function $\boldsymbol{V} \in \mathcal{U}$, compositely defined on suitable sets $\overline{\Omega}$ and $\Gamma$, is

$$\boldsymbol{V} := \Big(\boldsymbol{v}, z, \mathbf{N}, \mathbf{M}, \boldsymbol{f}, \boldsymbol{g}, \boldsymbol{n}\Big)^{\top}, \tag{1.4a}$$

where $\boldsymbol{v}$ and $\boldsymbol{f}$ are two dimensional vector fields and $z$ is a scalar field, that is,

$$\boldsymbol{v} : \overline{\Omega} \to \mathbb{R}^2, \quad \boldsymbol{f} : \overline{\Omega} \to \mathbb{R}^2, \quad z : \overline{\Omega} \to \mathbb{R}, \tag{1.4b}$$

$\mathbf{N}$ and $\mathbf{M}$ are symmetric tensor fields on $\mathbb{R}^2$ (hence identified with $\mathbb{R}^3$), and $\boldsymbol{g}$ and $\boldsymbol{n}$ are three dimensional vector fields, that is

$$\mathbf{N} : \overline{\Omega} \to \mathrm{Sym} \cong \mathbb{R}^3, \quad \mathbf{M} : \overline{\Omega} \to \mathrm{Sym} \cong \mathbb{R}^3, \quad \boldsymbol{g}, \boldsymbol{n} : \Gamma \to \mathbb{R}^3. \tag{1.4c}$$

Specifically, we essentially deal with a subspace of $\mathcal{U}$ (1.2), defined in the following:

$$\begin{aligned} \mathcal{D} := \{\boldsymbol{V} \in \mathcal{U} : \text{the composite structure (1.4) holds and} \\ (\mathtt{I} - \boldsymbol{e}_3 \otimes \boldsymbol{e}_3)\boldsymbol{n}' - \mathbf{N}\boldsymbol{l} = \mathbf{0} \text{ on } \Gamma, \\ \boldsymbol{n}' \cdot \boldsymbol{e}_3 - (\boldsymbol{f} \cdot \boldsymbol{l} + (\nabla \cdot \mathbf{M}) \cdot \boldsymbol{l} + \nabla(\boldsymbol{t} \cdot \mathbf{M}\boldsymbol{l}) \cdot \boldsymbol{t}) = 0 \text{ on } \Gamma, \\ \boldsymbol{l} \cdot \mathbf{M}\boldsymbol{l} = 0 \text{ on } \Gamma\}, \end{aligned} \tag{1.5}$$

where $\boldsymbol{l}$ and $\boldsymbol{t}$ are the outward unit normal on $\partial\Omega \equiv \Gamma$ and the unit tangent on (counter clockwise contour) $\Gamma$, respectively, $'$ refers to the derivative with respect to the arclength parameter of the curve $\Gamma$. Note that under the assumption of length scale (as mentioned in the context of (1.1)) $L = \frac{1}{2}\mathfrak{D}$, $\Gamma$ is simply the unit circle $\mathbb{T}$ in $\mathbb{R}^2$ so that $'$ refers the derivative along such unit circle.

Regarding the physical meaning of the quantities that enter in $\boldsymbol{V}$ defined above in (1.4a) and the conditions that appear in (1.5), we briefly digress in this paragraph; note that the details concerning the same are present in [5]. The constituents in (1.4a) are defined relative to the homogeneously deformed configuration. The first constituent in (1.4a), that is $\boldsymbol{v}$, denotes the in-plane displacement of the plate. The symbol $z$ in (1.4a) denotes the out-of-plane displacement of the plate. The symbols $\mathbf{N}$ and $\mathbf{M}$ in (1.4a) denote the membrane stress tensor (excluding stress in trivial solution) and bending couple tensor fields defined on the plate, respectively. The symbol $\boldsymbol{f}$ denotes the contribution of the membrane stress in the out-of-plane direction (hence related to $\mathbf{N}$ and $z$). The symbol $\boldsymbol{g}$ denotes the Gibbs rotation vector corresponding to the orientation of rod directors relative to that in the homogeneously deformed configuration. The symbol $\boldsymbol{n}$ denotes the internal forces in the rod that arise due to the inextensibility and unshearability constraints of the rod. The three conditions in (1.5) impose linear boundary conditions at the edge of plate. The first condition that appears in (1.5), i.e. $(\mathtt{I} - \boldsymbol{e}_3 \otimes \boldsymbol{e}_3)\boldsymbol{n}' - \mathbf{N}\boldsymbol{l} = \mathbf{0}$, describes the planar force equilibrium of the boundary, i.e., force equilibrium in radial and tangential directions. The second condition that appears in (1.5) is a scalar equation and encodes the force equilibrium of the boundary in out-of-plane direction. Finally, the third condition that appears in (1.5) prescribes the edge moment of the plate to be zero at the boundary, i.e., it represents the moment free condition at the plate boundary.

We also consider the space of functions denoted by

$$\mathcal{C} := (\mathcal{C}^0)^2 \times \mathcal{C}^0 \times (\mathcal{C}^0)^3 \times (\mathcal{C}^0)^3 \times (\mathcal{C}^1)^2 \times (\mathcal{P}^0)^3 \times (\mathcal{P}^0)^3, \tag{1.6}$$

using the acronyms in (1.3), where a typical element $\boldsymbol{Z} \in \mathcal{C}$ has the structure

$$\boldsymbol{Z} := \Big(\boldsymbol{u}, w, \mathbf{S}, \mathbf{T}, \boldsymbol{p}, \boldsymbol{q}, \boldsymbol{h}\Big)^\top. \tag{1.7}$$

In fact, the structure of (1.7) is reminiscent of (1.4a), and mainly we consider $\boldsymbol{Z}$ in a dense space of smooth functions same as in (1.4a). Thus, we assume that $\boldsymbol{u}$ and $\boldsymbol{p}$ are two dimensional vector fields, that is, $\boldsymbol{u}:\overline{\Omega}\to\mathbb{R}^2, \boldsymbol{p}:\overline{\Omega}\to\mathbb{R}^2$, $w$ is a scalar field, $w:\overline{\Omega}\to\mathbb{R}$, $\mathbf{S}$ and $\mathbf{T}$ are symmetric tensor fields (relative to the two dimensional plane $\mathbb{R}^2$), $\mathbf{S}:\overline{\Omega}\to \mathrm{Sym}\cong\mathbb{R}^3, \mathbf{T}:\overline{\Omega}\to \mathrm{Sym}\cong\mathbb{R}^3$, and $\boldsymbol{q}$, $\boldsymbol{h}$ are three dimensional vector fields, that is $\boldsymbol{g}:\Gamma\to\mathbb{R}^3, \boldsymbol{n}:\Gamma\to\mathbb{R}^3$.

The nonlinear problem of rod-plate equilibrium under perimeter mismatch is expressed in terms of an operator $\mathfrak{F}:\mathcal{D}\times\mathbb{R}^+\to\mathcal{C}$ defined by

$$\mathfrak{F}(\boldsymbol{V},\lambda):=\begin{pmatrix}\nabla\cdot\mathbf{N}\\ \nabla\cdot(\boldsymbol{f}+\nabla\cdot\mathbf{M})\\ 12\left((1-\nu)\mathbf{E}_\lambda+\nu(\mathrm{tr}\ \mathbf{E}_\lambda)\mathbf{I}\right)-\mathbf{N}\\ -\frac{h^2}{\lambda^2}((1-\nu)\nabla^2 z+\nu\Delta z\mathbf{I})-\mathbf{M}\\ \boldsymbol{f}-\frac{1}{\lambda}(12(1+\nu)(1-\lambda)\nabla z+\mathbf{N}\nabla z)\\ \boldsymbol{m}'-(\boldsymbol{n}+\frac{12}{\alpha}(1+\nu)(\lambda-1)\boldsymbol{t})\wedge(\boldsymbol{t}+\boldsymbol{v}'+z'\boldsymbol{e}_3)\\ \boldsymbol{t}+\boldsymbol{v}'+z'\boldsymbol{e}_3-\mathtt{R}(\boldsymbol{g})[\boldsymbol{t}]\end{pmatrix}\in\mathcal{C},\tag{1.8}$$

$$\boldsymbol{V}\in\mathcal{D},\quad \lambda\in\mathbb{R}^+,\quad \alpha:=2/\mathfrak{D},\tag{1.9}$$

where $\mathcal{D}$, domain of $\mathfrak{F}$, is defined by (1.5), $\mathcal{C}$, co-domain of $\mathfrak{F}$, is defined by (1.6),

$$\mathbf{E}_\lambda:=\tfrac{1}{2}(\nabla\boldsymbol{v}+\nabla\boldsymbol{v}^\top+\frac{1}{\lambda}\nabla z\otimes\nabla z),\quad \boldsymbol{m}:=\boldsymbol{m}_\perp+\boldsymbol{m}_{||},\tag{1.10a}$$

$$\boldsymbol{m}_\perp:=\beta\mathtt{R}(\boldsymbol{g})(\boldsymbol{t}\wedge\mathfrak{R}_{\boldsymbol{g}}\boldsymbol{t}),\quad \boldsymbol{m}_{||}:=\gamma(\boldsymbol{l}\cdot(\mathfrak{R}_{\boldsymbol{g}}\boldsymbol{e}_3]))\mathtt{R}(\boldsymbol{g})[\boldsymbol{t}],\tag{1.10b}$$

$$\mathfrak{R}_{\boldsymbol{g}}:=(\mathtt{R}(\boldsymbol{g}))^\top(\mathtt{R}(\boldsymbol{g}))',\ \mathtt{R}(\boldsymbol{g}):=\frac{((1-\boldsymbol{g}\cdot\boldsymbol{g})\mathtt{I}+2\boldsymbol{g}\otimes\boldsymbol{g}+2\mathrm{skw}(\boldsymbol{g}))}{1+\boldsymbol{g}\cdot\boldsymbol{g}},\tag{1.10c}$$

$$\text{and}\qquad \nu\in(0,\tfrac{1}{2}),\quad h>0,\quad \mathfrak{D}>0,\quad \beta>0,\quad \gamma>0.\tag{1.10d}$$

Eq. (1.10d) contains the five structure parameters of the rod-plate system, namely the plate material's Poisson ratio $\nu$ and the four structure parameters defined in (1.1).

*Remark* 1.1. As mentioned in the context of (1.1), under the assumption of length scale $L=\frac{1}{2}\overline{\mathfrak{D}}$, $\Gamma=\mathbb{T}\subset\mathbb{R}^2$ and $'$ in (1.8) refers to the derivative along $\mathbb{T}$ as a counter clockwise contour. As a consequence of the same length scale, also note that $\alpha=1$ in $(1.9)_3$, however, we retain the presence of $\alpha$ in most of the present article.

The nonlinear problem of rod-plate equilibrium is stated as

$$\mathfrak{F}(\boldsymbol{V},\lambda)=\mathbf{0},\quad \text{where } \boldsymbol{V}\in\mathcal{D} \text{ and } \lambda\in\mathbb{R}^+,\tag{1.11}$$

$\mathcal{D}$, the domain of $\mathfrak{F}$, is defined in (1.5), and $\mathfrak{F}$ itself is defined in (1.8)–(1.10). Regarding the domain $\mathcal{D}$ (1.2) of $\mathfrak{F}$, note that the solution of (1.11) belongs to a smoother class of functions; for an example to illustrate this, consider the function $z$, then it is clear that in place of $z\in\mathcal{C}^2$, actually $z\in\mathcal{C}^4$ by virtue of the fourth equation in the list (1.8) since $\mathbf{M}$ belongs to the class $\mathcal{C}^2$.

The derivation of nonlinear operator $\mathfrak{F}$ (1.11) is based on the physical problem of rod-plate equilibrium as mentioned in introduction; same is discussed with all details in [5]. However, to make this article self-contained, to some extent, we provide some further comments on the physical meaning of the constituents of (1.11), involving the operator $\mathfrak{F}$ in (1.8). The first constituent of (1.11), i.e. $\nabla\cdot\mathbf{N}=\mathbf{0}$, represents the equilibrium of planar stress in interior of the plate, given by divergence of membrane stress tensor. The second constituent describes the out-of-plane force balance in the interior of plate. The third and fourth constituents represent the constitutive relation for membrane stress and bending couple, respectively. The fifth constituent gives the expression for the contribution of the membrane stress on the out-of-plane force equilibrium. The last two terms serve as the nonlinear boundary conditions on the edge $\Gamma$, which is modelled as a rod. The sixth constituent represents the moment equilibrium of the rod. The seventh constituent gives the inextensibility and unshearability constraints combined into a single vector. The constituents of $\boldsymbol{Z}$ in (1.7) belong to the co-domain of the operator $\mathfrak{F}$ in (1.8). Some of the constituents in (1.7) lack a direct physical interpretation as they have been introduced purely for a operator-theoretic purpose; for example, a non-zero $\mathbf{S}$ or a non-zero $\boldsymbol{p}$ may correspond to a hypothetical constitutive relation/restriction.

For the purpose of the present article, specifically, it is given that $\mathfrak{F}(\mathbf{0}, \lambda) = \mathbf{0}$ for all $\lambda \in \mathbb{R}^+$. Thus, $\boldsymbol{V} = \mathbf{0}$ is the *trivial solution branch* that solves nonlinear problem (1.11) for all values of *bifurcation parameter* $\lambda$ (as shown in the extreme right of schematic Fig. 1). A natural question arises regarding nontrivial solution branches. The corresponding answer is provided by the local bifurcation analysis of solution set of (1.11) in a small neighbourhood of the trivial solution branch.

**2. Linearization.** The linear operator $\mathfrak{L}(\lambda) : \mathcal{D} \to \mathcal{C}$ obtained by linearizing $\mathfrak{F}$, defined in (1.8)–(1.10), about trivial solution branch, i.e. $\boldsymbol{V} = \mathbf{0}$ for all $\lambda \in \mathbb{R}^+$, along the vector $\boldsymbol{X} = \left(\boldsymbol{v}, z, \mathbf{N}, \mathbf{M}, \boldsymbol{f}, \mathbf{s}, \boldsymbol{n}\right)^\top \in \mathcal{D}$ is

$$\mathfrak{L}(\lambda)[\boldsymbol{X}] = \begin{pmatrix} \nabla \cdot \mathbf{N} \\ \nabla \cdot (\boldsymbol{f} + \nabla \cdot \mathbf{M}) \\ 12(\frac{(1-\nu)}{2}(\nabla \boldsymbol{v} + \nabla \boldsymbol{v}^\top) + \nu \nabla \cdot \boldsymbol{v}\mathbf{I}) - \mathbf{N} \\ -\frac{h^2}{\lambda^2}((1-\nu)\nabla^2 z + \nu \Delta z \mathbf{I}) - \mathbf{M} \\ \boldsymbol{f} - \frac{12}{\lambda}(1+\nu)(1-\lambda)\nabla z \\ 2\mathbf{C}[\mathbf{s}''] - 2\alpha(\gamma - \beta)(\boldsymbol{t} \otimes \boldsymbol{l} + \boldsymbol{l} \otimes \boldsymbol{t})[\mathbf{s}'] + \ldots \\ \ldots + \boldsymbol{t} \wedge (\boldsymbol{n} - \frac{12(1+\nu)(\lambda-1)}{\alpha}(\boldsymbol{v}' + z'\boldsymbol{e}_3)) \\ \boldsymbol{v}' + z'\boldsymbol{e}_3 - 2\mathbf{s} \wedge \boldsymbol{t} \end{pmatrix}, \mathbf{C} := \beta \mathtt{I} + (\gamma - \beta)\boldsymbol{t} \otimes \boldsymbol{t}, \tag{2.1}$$

where the vector field $\mathbf{s}$ arises from the linearization of the rotation tensor field, defined in (1.10c), that is $\mathtt{R}(\varepsilon \mathbf{s}) = \mathtt{I} + 2\varepsilon \operatorname{skw}(\mathbf{s}) + o(\varepsilon)$, as $\varepsilon \to 0$.

The derivation of the linear operator $\mathfrak{L}(\lambda)$, defined in (2.1), is rather routine. The auxiliary details concerning this are not included in this article. The relevant pieces of information from [5] are summarized in the following paragraph.

For each critical value $\lambda_c$, that is when the linear operator $\mathfrak{L}(\lambda)$ (2.1) is singular for $\lambda = \lambda_c$, the null space of $\mathfrak{L}(\lambda_c)$ can be found explicitly; see Appendix A for the form of null solutions. The adjoint operator $\mathfrak{L}^*$, corresponding to $\mathfrak{L}$, also plays an important role in the proceedings and is stated in (B.4); see Appendix B for related details. The set of all critical points $\{\lambda_c\} \subset \mathbb{R}^+$ can be sub-divided into different classes, for example, axisymmetric and non-axisymmetric, using symmetry associated with the null space corresponding to each critical point. It is found that the dimension of the null space for each critical point, *except* the axisymmetric ones, of the linear operator $\mathfrak{L}(\lambda_c)$ is 2. Hence, the sufficient condition of bifurcation does not hold [15] at those critical points where null solutions (vectors in the null space) are non-axisymmetric. However, there is a presence of symmetry in the nonlinear physical problem which can be used to reformulate the problem.

## 3. Symmetry-reduced formulation.

The problem admits a larger symmetry group $\mathtt{G}$ containing the aforementioned symmetries of the null spaces at all critical points. In fact, the null solutions at a given critical point remain unaltered for specific subgroup of symmetries of $\mathtt{G}$, that is they are stationary under specific subgroups of $\mathtt{G}$ known as isotropy sub-groups for those null solutions. The Theorem 2.3 of [11] is potentially application to obtain the sufficient conditions so that the null solutions of the linearized problem are guaranteed to be bifurcation points. Moreover, this idea can also assist in determination of the local bifurcation curves. It turns out, and what makes the analysis presented in this article successful, that the linear operator obtained by linearization about trivial solution for a suitable *symmetry-reduced formulation* of the nonlinear problem possesses a *one-dimensional* kernel even for those critical points corresponding to non-axisymmetric null solutions.

### 3.1. Symmetries of the problem.

The symmetries of the complete problem and null solutions are identified and expressed as groups in this part of the section. In particular, the generating set of the complete symmetry group has three elements $r_\theta, e$, and $f$, which are described in the following list:

- $r_\theta$: counter clockwise rotation by angle $\theta$ about axis passing through the centre of the plate and perpendicular to the plane of the plate.
- $e$: reflection about a plane perpendicular to the plate containing the diameter parallel to $\boldsymbol{e}_1$ axis.
- $f$: 180° flip about diameter parallel to $\boldsymbol{e}_1$ axis.

The complete symmetry group is represented by

$$\mathtt{G} := \{r_\theta, er_\theta, fr_\theta, fer_\theta \colon \theta \in \mathbb{R}_{2\pi}\}, \quad \text{where } r_{2\pi} = \iota, \tag{3.1a}$$

$$e, f \text{ satisfy } e^2 = \iota, f^2 = \iota, \qquad efef = \iota, \qquad er_\theta er_\theta = \iota, fr_\theta fr_\theta = \iota, \tag{3.1b}$$

and $\iota$ is the identity element. The set notation used in (3.1a) considers $\theta$ as a continuous parameter spanning over elements of group.

*Remark* 3.1. The complete symmetry group $\mathtt{G}$ as stated above contains all those symmetries of the trivial solution branch which are either broken or retained by the null solutions [5]. There may be other symmetries in the nonlinear operator $\mathfrak{F}$, outside $\mathtt{G}$, but they are not broken at any critical point and hence not discussed in this article.

The subgroups of the complete symmetry group $\mathtt{G}$, that are relevant for the concerned rod-plate problem, are

$$\mathtt{D}_n := \{r_\zeta, r_{2\zeta}, \ldots, r_{n\zeta}, er_\zeta, er_{2\zeta}, \ldots, er_{n\zeta}\}, \quad \text{where } \zeta = 2\pi/n; \tag{3.2a}$$

$$\begin{aligned}\mathtt{Z}_n := \{&r_\zeta, r_{2\zeta}, \ldots, r_{n\zeta}, er_\zeta, er_{2\zeta}, \ldots, er_{n\zeta},\\ &fr_\zeta, fr_{2\zeta}, \ldots, fr_{n\zeta}, fer_\zeta, fer_{2\zeta}, \ldots, fer_{n\zeta}\}, \quad \text{where } \zeta = 2\pi/n;\end{aligned} \tag{3.2b}$$

$$\mathtt{O}(2) := \{r_\theta, er_\theta \colon \theta \in \mathbb{R}_{2\pi}\}, \text{where} \quad r_{2\pi} = \iota. \tag{3.2c}$$

These three subgroups $\mathtt{D}_n$, $\mathtt{Z}_n$, and $\mathtt{O}(2)$ (defined in (3.2a), (3.2b), and (3.2c), respectively) of the complete symmetry group $\mathtt{G}$ are used to find the symmetry-reduced function spaces for each such subgroup.

**3.2. Symmetry-reduced formulation of the nonlinear operator.** We consider the following linear transformations to construct the symmetry-reduced spaces corresponding to the three subgroups $\mathtt{D}_n$, $\mathtt{Z}_n$, and $\mathtt{O}(2)$, defined in (3.2a), (3.2b), and (3.2c), of the complete symmetry group $\mathtt{G}$:

$$\mathbf{J} : \mathbb{R}^2 \to \mathbb{R}^2, \quad \mathbf{J} := \boldsymbol{e}_1 \otimes \boldsymbol{e}_1 - \boldsymbol{e}_2 \otimes \boldsymbol{e}_2 \ \ \text{(reflection about } \boldsymbol{e}_1 \text{ axis)}; \tag{3.3a}$$

$$\mathbf{T}_\phi : \mathbb{R}^2 \to \mathbb{R}^2, \ \mathbf{T}_\phi := \cos\phi\, (\boldsymbol{e}_1 \otimes \boldsymbol{e}_1 + \boldsymbol{e}_2 \otimes \boldsymbol{e}_2) + \sin\phi\, (\boldsymbol{e}_2 \otimes \boldsymbol{e}_1 - \boldsymbol{e}_1 \otimes \boldsymbol{e}_2) \tag{3.3b}$$

(counter clockwise rotation by an angle $\phi$);

$$\mathtt{Q} : \mathbb{R}^3 \to \mathbb{R}^3, \ \mathtt{Q} := \boldsymbol{e}_1 \otimes \boldsymbol{e}_1 - \boldsymbol{e}_2 \otimes \boldsymbol{e}_2 + \boldsymbol{e}_3 \otimes \boldsymbol{e}_3 \ \text{(reflection about } \boldsymbol{e}_1\text{-}\boldsymbol{e}_3 \text{ plane)}; \tag{3.3c}$$

$$\mathtt{F} : \mathbb{R}^3 \to \mathbb{R}^3, \quad \mathtt{F} := \boldsymbol{e}_1 \otimes \boldsymbol{e}_1 - \boldsymbol{e}_2 \otimes \boldsymbol{e}_2 - \boldsymbol{e}_3 \otimes \boldsymbol{e}_3 \ \ \text{(flip about } \boldsymbol{e}_1 \text{ axis)}; \tag{3.3d}$$

$$\mathtt{Z}_\phi : \mathbb{R}^3 \to \mathbb{R}^3, \quad \mathtt{Z}_\phi := \cos\phi\, (\boldsymbol{e}_1 \otimes \boldsymbol{e}_1 + \boldsymbol{e}_2 \otimes \boldsymbol{e}_2) + \sin\phi\, (\boldsymbol{e}_2 \otimes \boldsymbol{e}_1 - \boldsymbol{e}_1 \otimes \boldsymbol{e}_2) + \boldsymbol{e}_3 \otimes \boldsymbol{e}_3 \ \ \text{(rotation about } \boldsymbol{e}_3 \text{ axis by an angle } \phi). \tag{3.3e}$$

We use the polar coordinates $(r, \theta)$ to represent a point in $\overline{\Omega}$. Let

$$\boldsymbol{e}_r := \cos\theta \boldsymbol{e}_1 + \sin\theta \boldsymbol{e}_2, \quad \boldsymbol{e}_\theta := -\sin\theta \boldsymbol{e}_1 + \cos\theta \boldsymbol{e}_2, \tag{3.4}$$

where $\boldsymbol{e}_r, \boldsymbol{e}_\theta$ correspond to the usual polar basis vectors. In this context, it is useful to recall Remark 1.1.

The operator $\mathfrak{F}$ in (1.8) is said to be equivariant under symmetry transformation $\mathcal{S}$ if

$$\mathfrak{F}(\mathcal{S}\boldsymbol{V}, \lambda) = \mathcal{S}(\mathfrak{F}(\boldsymbol{V}, \lambda)), \ \forall \boldsymbol{V} \in \mathcal{D}, \lambda \in \mathbb{R}^+. \tag{3.5}$$

In particular, by restricting the above condition to the generators of $\mathtt{G}$, the operator $\mathfrak{F}$ in (1.8) is equivariant under symmetry $\mathtt{G}$ if, for an element, according to the definition (1.4a), in the space $\mathcal{D}$, we have

$$\mathfrak{F}(\mathcal{T}_\phi\boldsymbol{V}, \lambda) = \mathcal{T}_\phi(\mathfrak{F}(\boldsymbol{V}, \lambda)), \mathfrak{F}(\mathcal{E}\boldsymbol{V}, \lambda) = \mathcal{E}(\mathfrak{F}(\boldsymbol{V}, \lambda)), \mathfrak{F}(\mathcal{F}\boldsymbol{V}, \lambda) = \mathcal{F}(\mathfrak{F}(\boldsymbol{V}, \lambda)), \tag{3.6}$$

$\forall \boldsymbol{V} \in \mathcal{D}, \lambda \in \mathbb{R}^+$, where $\mathcal{T}_\phi$, $\mathcal{E}$ and $\mathcal{F}$ are the symmetry transformations corresponding to rotation $(r_\phi)$, reflection $(e)$ and flip $(f)$, respectively, and given by

$$(\mathcal{T}_\phi\boldsymbol{V})(r,\theta) = \Big(\mathbf{T}_\phi\boldsymbol{v}, z, \mathbf{T}_\phi\mathbf{N}\mathbf{T}_\phi^\top, \mathbf{T}_\phi\mathbf{M}\mathcal{T}_\phi^\top, \mathbf{T}_\phi\boldsymbol{f}, \mathtt{Z}_\phi\boldsymbol{g}, \mathtt{Z}_\phi\boldsymbol{n}\Big)^\top (r, \theta-\phi), \tag{3.7a}$$

$$(\mathcal{E}\boldsymbol{V})(r,\theta) = \Big(\mathbf{J}\boldsymbol{v}, z, \mathbf{JNJ}, \mathbf{JMJ}, \mathbf{J}\boldsymbol{f}, -\mathtt{Q}\boldsymbol{g}, -\mathtt{Q}\boldsymbol{n}\Big)^\top (r, -\theta), \tag{3.7b}$$

$$(\mathcal{F}\boldsymbol{V})(r,\theta) = \Big(\mathbf{J}\boldsymbol{v}, -z, \mathbf{JNJ}, -\mathbf{JMJ}, -\mathbf{J}\boldsymbol{f}, \mathtt{F}\boldsymbol{g}, -\mathtt{F}\boldsymbol{n}\Big)^\top (r, -\theta). \tag{3.7c}$$

That the operator $\mathfrak{F}$ in (1.8) is equivariant under the action of $\mathtt{G}$ involves some lengthy calculations but these are rather elementary; the details regarding the proof of the equivariance of nonlinear operator the action of $\mathtt{G}$ are provided in supplementary S1.

As the operator $\mathfrak{F}$ in (1.8) is equivariant under the action of $\mathtt{G}$, the reduction can be done simply by restricting the associated function spaces using the projections constructed in the sequel. Recall that $\mathcal{D}$ is the domain of $\mathfrak{F}$, defined by (1.5), and $\mathcal{C}$ is the co-domain of $\mathfrak{F}$, defined by (1.6). In order to construct the projections for a symmetry-reduced formulation, using (3.3a)–(3.3e), the following linear transforma-tions are defined on $\mathcal{D}$ and $\mathcal{C}$:

$$\mathbf{G}_F[\boldsymbol{V}(r,\theta)] := \Big(\mathbf{J}\boldsymbol{v}_1, -v_2, \mathbf{J}\mathbf{V}_3\mathbf{J}, -\mathbf{J}\mathbf{V}_4\mathbf{J}, -\mathbf{J}\boldsymbol{v}_5, \mathtt{F}\boldsymbol{v}_6, -\mathtt{F}\boldsymbol{v}_7\Big)^\top(r,\theta), \tag{3.8a}$$

which describes the flip transformation about $\boldsymbol{e}_1$-axis;

$$\mathbf{G}_E[\boldsymbol{V}(r,\theta)] := \Big(\mathbf{J}\boldsymbol{v}_1, v_2, \mathbf{J}\mathbf{V}_3\mathbf{J}, \mathbf{J}\mathbf{V}_4\mathbf{J}, \mathbf{J}\boldsymbol{v}_5, -\mathtt{Q}\boldsymbol{v}_6, -\mathtt{Q}\boldsymbol{v}_7\Big)^\top(r,\theta), \tag{3.8b}$$

which describes the reflection about $\boldsymbol{e}_1$-$\boldsymbol{e}_3$ plane;

$$\mathbf{G}_\phi[\boldsymbol{V}(r,\theta)] := \Big(\mathbf{T}_\phi\boldsymbol{v}_1, v_2, \mathbf{T}_\phi\mathbf{V}_3\mathbf{T}_\phi^\top, \mathbf{T}_\phi\mathbf{V}_4\mathbf{T}_\phi^\top, \mathbf{T}_\phi\boldsymbol{v}_5, \mathtt{Z}_\phi\boldsymbol{v}_6, \mathtt{Z}_\phi\boldsymbol{v}_7\Big)^\top(r,\theta), \tag{3.8c}$$

which describes the (counter clockwise) rotation by an angle $\phi$ about $\boldsymbol{e}_3$-axis.

In connection with (3.8) and the structure in the definition (1.4a), we use the notation

$$\boldsymbol{V} = \Big(\boldsymbol{v}_1, v_2, \mathbf{V}_3, \mathbf{V}_4, \boldsymbol{v}_5, \boldsymbol{v}_6, \boldsymbol{v}_7\Big)^\top. \tag{3.9}$$

*Remark* 3.2. In general, the notational scheme (3.9) can be stated as follows. We use the short hand notation $(a_1, \dots, a_7)(r,\theta)$ to represent $(a_1(r,\theta), \dots, a_5(r,\theta),\ a_6(\theta), \dots, a_7(\theta))$ where $a_1, \dots a_7$ are according to the structure in definition (1.4a).

**3.2.1. $\mathtt{D}_n$ symmetry.** The action of $\mathtt{D}_n$ on an element, according to the definition (1.4a), in the space $\mathcal{D}$ is:

$$(\mathtt{D}_n^{(j)}\boldsymbol{V})(r,\theta) := \begin{cases}\mathbf{G}_{j\zeta}[\boldsymbol{V}(r,\theta-j\zeta)] & \forall j\in\{1,2,\dots,n\}\\ \mathbf{G}_E[\mathbf{G}_{j\zeta}[\boldsymbol{V}(r,-\theta-j\zeta)]] & \forall j\in\{1,2,\dots,n\},\quad \zeta=2\pi/n\end{cases}. \tag{3.10}$$

The projection $\mathcal{T}_{\mathtt{D}_n}: \mathcal{D}\to\mathcal{D}_{\mathtt{D}_n}$ is given by (Eq. 3.29 of [33])

$$(\mathcal{T}_{\mathtt{D}_n}\boldsymbol{V})(r,\theta) = \tfrac{1}{2n}\sum\nolimits_{j=1}^{n}\left(\mathbf{G}_{j\zeta}[\boldsymbol{V}(r,\theta-j\zeta)]\right) + \tfrac{1}{2n}\sum\nolimits_{j=1}^{n}\left(\mathbf{G}_E[\mathbf{G}_{j\zeta}[\boldsymbol{V}(r,-\theta-j\zeta)]]\right). \tag{3.11}$$

Using polar coordinates and (3.4) and $(1.4c)_3$, in connection with $(1.4b)_{1,2}$,

$$\begin{aligned}\boldsymbol{v} = v_r\boldsymbol{e}_r + v_\theta\boldsymbol{e}_\theta, \quad \boldsymbol{f} = f_r\boldsymbol{e}_r + f_\theta\boldsymbol{e}_\theta, \quad \boldsymbol{g} &= g_r\boldsymbol{e}_r + g_\theta\boldsymbol{e}_\theta + g_3\boldsymbol{e}_3,\\ \boldsymbol{n} &= n_r\boldsymbol{e}_r + n_\theta\boldsymbol{e}_\theta + n_3\boldsymbol{e}_3,\end{aligned} \tag{3.12}$$

and the tensors $\mathbf{N}$ and $\mathbf{M}$ $(1.4c)_{1,2}$ are expressed, in (polar) components, as

$$\mathbf{N} = N_{rr}\boldsymbol{e}_r\otimes\boldsymbol{e}_r + N_{\theta\theta}\boldsymbol{e}_\theta\otimes\boldsymbol{e}_\theta + N_{r\theta}(\boldsymbol{e}_r\otimes\boldsymbol{e}_\theta + \boldsymbol{e}_\theta\otimes\boldsymbol{e}_r), \tag{3.13}$$

$$\mathbf{M} = M_{rr}\boldsymbol{e}_r\otimes\boldsymbol{e}_r + M_{\theta\theta}\boldsymbol{e}_\theta\otimes\boldsymbol{e}_\theta + M_{r\theta}(\boldsymbol{e}_r\otimes\boldsymbol{e}_\theta + \boldsymbol{e}_\theta\otimes\boldsymbol{e}_r). \tag{3.14}$$

In the context of the definition of $\boldsymbol{V}$ (1.4a), we identify $\boldsymbol{v}$ with $(v_r, v_\theta)$, $\boldsymbol{f}$ with $(f_r, f_\theta)$, $\boldsymbol{g}$ with $(g_r, g_\theta, g_3)$, $\boldsymbol{n}$ with $(n_r, n_\theta, n_3)$, and in accordance with $(1.4c)_{1,2}$, we identify the tensor $\mathbf{N}$ with the triplet $[N_{rr}, N_{\theta\theta}, N_{r\theta}]$ and $\mathbf{M}$ with the triplet $[M_{rr}, M_{\theta\theta}, M_{r\theta}]$.

*Remark* 3.3. The notation described in Remark 3.2 is used even when the individual elements $a_i, i = 1, \dots, 7$ in $(a_1, \dots, a_7)(r,\theta)$ are written in suitable components. For example, we use the succint notation $\boldsymbol{V}(r,\theta)$ to replace the long expression $\Big((v_r, v_\theta), z, [N_{rr}, N_{\theta\theta}, N_{r\theta}], [M_{rr}, M_{\theta\theta}, M_{r\theta}], (f_r, f_\theta), (g_r, g_\theta, g_3), (n_r, n_\theta, n_3)\Big)^\top$, whenever it is convenient and unambigious.

With above provisions for the components, substituting (1.4a) in (3.11), and expanding (recall Remark 3.2, 3.3), we get

$$\begin{aligned}(\mathcal{T}_{\mathtt{D}_n}\boldsymbol{V})(r,\theta) = \tfrac{1}{2n}\sum_{j=1}^{n} \boldsymbol{V}(r,\theta - j\zeta) \\ + \tfrac{1}{2n}\sum_{j=1}^{n} \Big( (v_r, -v_\theta), z, \big[N_{rr}, N_{\theta\theta}, -N_{r\theta}\big], \big[M_{rr}, M_{\theta\theta}, -M_{r\theta}\big], \\ (f_r, -f_\theta), \big( -g_r, g_\theta, -g_3\big), \big( -n_r, n_\theta, -n_3\big)\Big)^{\top}(r, -\theta - j\zeta).\end{aligned} \tag{3.15}$$

For $\boldsymbol{V}$ to be stationary under $\mathtt{D}_n$, the following condition (obtained by using Eq. (2.1) of [11] with the defining expression of the projection $\mathcal{T}_{\mathtt{D}_n}$ (3.11)) must hold:

$$(\mathcal{T}_{\mathtt{D}_n}\boldsymbol{V})(r,\theta) = \boldsymbol{V}(r,\theta). \tag{3.16}$$

Due to (3.15) and (3.16), a necessary condition is that the components of $\boldsymbol{V}$ have period $2\pi/n$ in $\theta$. Applying the periodic condition to (3.15) and substituting (3.15) in (3.16), the restriction on reduced space simplifies to

$$\begin{aligned}\boldsymbol{V}(r,\theta) = \Big( (v_r, -v_\theta), z, \big[N_{rr}, N_{\theta\theta}, -N_{r\theta}\big], \big[M_{rr}, M_{\theta\theta}, -M_{r\theta}\big], (f_r, -f_\theta), \\ \big( -g_r, g_\theta, -g_3\big), \big( -n_r, n_\theta, -n_3\big)\Big)^{\top}(r, -\theta).\end{aligned} \tag{3.17}$$

**3.2.2. $\mathtt{Z}_n$ symmetry.** With $\zeta = 2\pi/n$, action of $\mathtt{Z}_n$ on an element, according to the definition (1.4a), of space $\mathcal{D}$ is

$$(\mathtt{Z}_n^{(j)}\boldsymbol{V})(r,\theta) = \begin{cases} \mathbf{G}_{j\zeta}[\boldsymbol{V}(r,\theta - j\zeta)] & \forall j \in \{1,2,\dots,n\} \\ \mathbf{G}_{E}[\mathbf{G}_{j\zeta}[\boldsymbol{V}(r,-\theta - j\zeta)]] & \forall j \in \{1,2,\dots,n\} \\ \mathbf{G}_{F}[\mathbf{G}_{j\zeta}[\boldsymbol{V}(r,-\theta - j\zeta)]] & \forall j \in \{1,2,\dots,n\} \\ \mathbf{G}_{F}[\mathbf{G}_{E}[\mathbf{G}_{j\zeta}[\boldsymbol{V}(r,\theta - j\zeta)]]] & \forall j \in \{1,2,\dots,n\}. \end{cases} \tag{3.18}$$

With above provisions for the components same as above in §3.2.1, similar to (3.15), the projection $\mathcal{T}_{\mathtt{Z}_n} : \mathcal{D} \to \mathcal{D}_{\mathtt{Z}_n}$ is given by (recall Remark 3.2, 3.3)

$$\begin{aligned}(\mathcal{T}_{\mathtt{Z}_n}\boldsymbol{V})(r,\theta) =& \tfrac{1}{2n}\sum_{j=1}^{n} \Big( (v_r, v_\theta), 0, \big[N_{rr}, N_{\theta\theta}, N_{r\theta}\big], \big[0,0,0\big], (0,0), \\ & (0,0,g_3), (n_r, n_\theta, 0)\Big)^{\top}(r, \theta - j\zeta) \\ & + \tfrac{1}{2n}\sum_{j=1}^{n} \Big( (v_r, -v_\theta), 0, \big[N_{rr}, N_{\theta\theta}, -N_{r\theta}\big], \big[0,0,0\big], (0,0), \\ & (0,0,-g_3), \big( -n_r, n_\theta, 0\big)\Big)^{\top}(r, -\theta - j\zeta).\end{aligned} \tag{3.19}$$

Also, in a manner similar to (3.16), for $\boldsymbol{V}$ to be stationary under $\mathtt{Z}_n$, we require that

$$(\mathcal{T}_{\mathtt{Z}_n}\boldsymbol{V})(r,\theta) = \boldsymbol{V}(r,\theta). \tag{3.20}$$

A necessary condition, for (3.19) and (3.20) to hold, is that the polar components of $\boldsymbol{V}$ have period $2\pi/n$ in $\theta$. Applying this periodic condition to (3.19) and substituting (3.19) in (3.20), the restriction on reduced space simplifies to (recall Remark 3.2, 3.3)

$$\begin{aligned}\boldsymbol{V}(r,\theta) =& \Big( (v_r, -v_\theta), 0, \big[N_{rr}, N_{\theta\theta}, -N_{r\theta}\big], \big[0,0,0\big], (0,0), \\ & (0,0,-g_3), \big( -n_r, n_\theta, 0\big)\Big)^{\top}(r, -\theta).\end{aligned} \tag{3.21}$$

**3.2.3. $\mathtt{O}(2)$ symmetry.** The action of $\mathtt{O}(2)$ on an element, according to the definition (1.4a), of space $\mathcal{D}$ is given by

$$(\mathtt{O}(2)\boldsymbol{V})(r,\theta)=\begin{cases}\mathbf{G}_\phi[\boldsymbol{V}(r,\theta-\phi)] & \forall\phi\in\mathbb{R}_{2\pi},\\ \mathbf{G}_E[\mathbf{G}_\phi[\boldsymbol{V}(r,-\theta-\phi)]] & \forall\phi\in\mathbb{R}_{2\pi}.\end{cases} \tag{3.22}$$

The projection $\mathcal{T}_{\mathtt{O}(2)}:\mathcal{D}\to\mathcal{D}_{\mathtt{O}(2)}$ is given by (see Eq. (3.30) of [33]),

$$(\mathcal{T}_{\mathtt{O}(2)}\boldsymbol{V})(r,\theta)=\tfrac{1}{4\pi}\int_0^{2\pi}(\mathbf{G}_\phi[\boldsymbol{V}(r,\theta-\phi)])\,\mathrm{d}\phi+\tfrac{1}{4\pi}\int_0^{2\pi}(\mathbf{G}_E[\mathbf{G}_\phi[\boldsymbol{V}(r,-\theta-\phi)]])\,\mathrm{d}\phi, \tag{3.23}$$

which can be expanded as (recall Remark 3.2, 3.3)

$$\begin{aligned}(\mathcal{T}_{\mathtt{O}(2)}\boldsymbol{V})(r,\theta)&=\tfrac{1}{4\pi}\int_0^{2\pi}\boldsymbol{V}(r,\theta-\phi)\mathrm{d}\phi\\&+\tfrac{1}{4\pi}\int_0^{2\pi}\Big((v_r,-v_\theta),z,\big[N_{rr},N_{\theta\theta},-N_{r\theta}\big],\big[M_{rr},M_{\theta\theta},-M_{r\theta}\big],\\&\big(f_r,-f_\theta\big),\big(-g_r,g_\theta,-g_3\big),\big(-n_r,n_\theta,-n_3\big)\Big)^\top(r,-\theta-\phi)\mathrm{d}\phi.\end{aligned} \tag{3.24}$$

Similar to (3.16), for $\boldsymbol{V}$ to be stationary under $\mathtt{O}(2)$, it must satisfy the condition

$$(\mathcal{T}_{\mathtt{O}(2)}\boldsymbol{V})(r,\theta)=\boldsymbol{V}(r,\theta). \tag{3.25}$$

As components of $\boldsymbol{V}$ are $2\pi$-periodic in $\theta$, the integration in (3.24) involving the $\theta$ dependent terms vanishes. Hence, (3.25) implies that $\boldsymbol{V}$ is only a function of $r$. Thus, we obtain the following form of $\boldsymbol{V}$ that is stationary under $\mathtt{O}(2)$ symmetry:

$$\boldsymbol{V}(r,\theta)=\Big((v_r,0),z,\big[N_{rr},N_{\theta\theta},0\big],\big[M_{rr},M_{\theta\theta},0\big],\big(f_r,0\big),\big(0,g_\theta,0\big),\big(0,n_\theta,0\big)\Big)^\top(r). \tag{3.26}$$

**3.3. Sufficient conditions for bifurcation.** Let us recall Theorem 2.3 (portion dealing with local analysis only) of [11] that gives sufficient conditions for a critical point to be a bifurcation point . Rewriting the same, we have

THEOREM 3.4. *Suppose that $\mathfrak{F}$ is twice continuously differentiable, Fredholm operator, and that there exists a null solution $\boldsymbol{X}\in\mathcal{N}(\mathfrak{L}(\lambda_c))$ that defines proper isotropy subgroup* $\mathtt{H}$. *Let $\mathfrak{L}_{\mathtt{H}}(\lambda_c)$ be the* $\mathtt{H}$*-reduced linearized operator about solution $(\mathbf{0},\lambda_c)$. Assume:*

*(1)* $\dim(\mathcal{N}(\mathfrak{L}_{\mathtt{H}}(\lambda_c)))$ *is odd.*

*(2)* $\mathrm{D}_\lambda\mathfrak{L}_{\mathtt{H}}(\lambda_c)[\boldsymbol{X}]\notin\mathcal{R}(\mathfrak{L}_{\mathtt{H}}(\lambda_c))\quad\forall\boldsymbol{X}\in\mathcal{N}(\mathfrak{L}_{\mathtt{H}}(\lambda_c))$

*Then, $(\mathbf{0},\lambda_c)$ is a bifurcation point of (1.8) such that in every small neighbourhood of $(\mathbf{0},\lambda_c)$, there are nontrivial solution $(\boldsymbol{V}_*,\lambda_*)\in\mathcal{D}_{\mathtt{H}}\times\mathbb{R}^+$. In particular if* $\dim(\mathcal{N}(\mathfrak{L}_{\mathtt{H}}(\lambda_c)))=1$, *then there exist a unique, local, bifurcating branch of solution of form $t\mapsto(\boldsymbol{V}(t),\lambda(t))\in\mathcal{D}_{\mathtt{H}}\times\mathbb{R}^+$.*

The relevant $\mathtt{H}$-reduced entities are discussed in the following paragraph.

We define $\mathtt{H}$-reduced operator $\mathfrak{F}_{\mathtt{H}}:\mathcal{D}_{\mathtt{H}}\times\mathbb{R}^+\to\mathcal{C}_{\mathtt{H}}$, as follows:

$$\mathfrak{F}_{\mathtt{H}}(\boldsymbol{u},\lambda):=\mathcal{T}_{\mathtt{H}}\mathfrak{F}(\boldsymbol{u},\lambda),\quad\forall\boldsymbol{u}\in\mathcal{D}_{\mathtt{H}},\ \lambda\in\mathbb{R}^+, \tag{3.27}$$

where the $\mathtt{H}$-reduced function spaces are

$$\mathcal{D}_{\mathtt{H}}:=\{\boldsymbol{u}\in\mathcal{D}:\mathbf{G}_{\mathrm{h}}\boldsymbol{u}=\boldsymbol{u},\ \forall h\in\mathtt{H}\}, \tag{3.28}$$

$$\mathcal{C}_{\mathtt{H}}:=\{\boldsymbol{u}\in\mathcal{C}:\mathbf{G}_{\mathrm{h}}\boldsymbol{u}=\boldsymbol{u},\ \forall h\in\mathtt{H}\}, \tag{3.29}$$

and $\mathcal{T}_{\mathtt{H}}$ is a projection from $\mathcal{D}\to\mathcal{D}_{\mathtt{H}}$, and $\mathcal{C}\to\mathcal{C}_{\mathtt{H}}$.

*Remark* 3.5. In order to apply Theorem 3.4, we use the fact that the linear operator $\mathfrak{L}$ is a Fredholm operator, based on arguments and definitions of domain (in place of $\mathcal{D}$ in (3.28) for $\mathcal{D}_{\mathtt{H}}$) and codomain (in place of $\mathcal{C}$ in (3.29) for $\mathcal{C}_{\mathtt{H}}$) similar to those provided in §III.1 of [15] for elliptic partial differential equations. A similar set of arguments hold regarding the (twice) Fréchet differentiability of $\mathfrak{F}$, using the same definitions of domain and codomain.

The reducing projections mapping the original function space on to symmetry-reduced function space are constructed above in (3.17), (3.21), and (3.26), corresponding to $\mathtt{D}_n$, $\mathtt{Z}_n$ and $\mathtt{O}(2)$ symmetry groups, respectively. These symmetry groups are the isotropy subgroups corresponding to the suitable null solutions of the linearized problem. As the problem is equivariant under the actions of the stated symmetries, the reduced operator is obtained by restricting the domain and codomain spaces. Note that

$$\mathcal{T}_{\mathtt{H}}\mathfrak{F}(\boldsymbol{u},\lambda) = \mathfrak{F}(\mathcal{T}_{\mathtt{H}}\boldsymbol{u},\lambda) = \mathfrak{F}(\boldsymbol{u},\lambda), \quad \forall \boldsymbol{u} \in \mathcal{D}_{\mathtt{H}},\ \lambda \in \mathbb{R}^+, \tag{3.30}$$

where $\mathtt{H}$ is one of the three symmetry groups, namely $\mathtt{D}_n$, $\mathtt{Z}_n$, and $\mathtt{O}(2)$, noted earlier corresponding to the critical points $\lambda_c$. The first equality in (3.30) is due to the equivariance of $\mathfrak{F}$ and the second equality is due to the projection $\mathcal{T}_{\mathtt{H}}$ acting on an element of $\mathcal{D}_{\mathtt{H}}$. Using the $\mathtt{H}$-reduced versions of $\mathcal{D}$ and $\mathcal{C}$, consequently, we also obtain the $\mathtt{H}$-reduced operators corresponding to $\mathfrak{L}$ (2.1) and $\mathfrak{L}^*$ (B.4), which are denoted by $\mathfrak{L}_{\mathtt{H}}$ and $\mathfrak{L}^*_{\mathtt{H}}$, respectively.

As mentioned before, for each critical value $\lambda = \lambda_c$ of the linear operator $\mathfrak{L}(\lambda)$ (2.1), the form of the null space of $\mathfrak{L}(\lambda_c)$ is stated in Appendix A. The symmetries of null spaces have been also stated and described in detail in [5].

Consider first the case of critical values which correspond to a non-planar, non-axisymmetric null solution of $\mathfrak{L}(\lambda_c)$. It can be seen that only $\boldsymbol{X}_1$ of (A.1) satisfies (3.17). Thus, the $\mathtt{D}_n$ reduced problem has a one dimensional kernel at critical points with non-planar null solution. Hence, the condition (1) of Theorem 3.4 is satisfied for the $\mathtt{D}_n$ reduced problem at such critical points.

Secondly, consider the case of critical values which correspond to a planar, non-axisymmetric null solution of $\mathfrak{L}(\lambda_c)$. It can be seen that only $\boldsymbol{X}_1$ of (A.2) satisfies (3.21). Thus, the $\mathtt{Z}_n$ reduced problem has a one dimensional kernel at critical points with planar null solution. Hence, the condition (1) of Theorem 3.4 is also satisfied for the $\mathtt{Z}_n$ reduced problem at such critical points.

That is, in summary of first and second cases, $\mathtt{D}_n$ is the isotropy subgroup for one of the non-planar null solutions of $\mathfrak{L}(\lambda_c)$, while $\mathtt{Z}_n$, the union of $\mathtt{D}_n$ and flip about a diameter, gives the isotropy group for one of the planar null solutions.

Finally, consider the case of critical values which correspond to a axisymmetric null solution of $\mathfrak{L}(\lambda_c)$. The null solutions of $\mathfrak{L}(\lambda_c)$ for the axisymmetric case are static under the action of orthogonal group $\mathtt{O}(2)$. Such critical values have already one dimensional null space (see (A.3) in Appendix A).

Thus, the condition (1) of Theorem 3.4 holds since $\dim \mathcal{N}(\mathfrak{L}_{\mathtt{H}}(\lambda_c)) = 1$ for all critical values $\lambda_c$.

In the remainder of the article, after checking the transversality property (condition (2) of Theorem 3.4), we carry out the Lyapunov–Schmidt reduction on the symmetry-reduced problem, corresponding to $\mathtt{D}_n$, $\mathtt{Z}_n$ and $\mathtt{O}(2)$ symmetry groups, to obtain the bifurcation curves up to quadratic order.

## 4. Local bifurcation analysis.

**4.1. Transversality condition.** The second condition in Theorem 3.4 is rewritten as a calculable expression involving null solution of $\mathfrak{L}^*_{\mathtt{H}}$, the adjoint of the linear operator $\mathfrak{L}_{\mathtt{H}}$. The same emphasises the need for finding expression of the adjoint operator (as already mentioned in §3), which is found to be formally same as that stated as $\mathfrak{L}^*$ (B.4) (as provided in Appendix B) for verification of transversality condition.

Further, we use the property that $\langle \boldsymbol{Y}, \boldsymbol{Z}\rangle = 0, \forall \boldsymbol{Y} \in \mathcal{R}(\mathfrak{L})$, for general linear operator $\mathfrak{L}$. The equivalent property to the condition (2) of Theorem 3.4, in the context of the linearized operator $\mathfrak{L}_{\mathtt{H}}(\lambda_c)$, at the critical point $\lambda_c$, is that $\langle \boldsymbol{Y}, \boldsymbol{Z}\rangle \neq 0 \implies \boldsymbol{Y} \notin \mathcal{R}(\mathfrak{L}_{\mathtt{H}}(\lambda_c))$, where $\boldsymbol{Z}$ is a null solution of the reduced adjoint operator and $\langle \cdot, \cdot\rangle$ is a suitable inner product. The details are presented in the next section. Consequently, the condition (2) of Theorem 3.4 can be re-written as

$$\langle \mathrm{D}_\lambda \mathfrak{L}_{\mathtt{H}}(\lambda_c)[\boldsymbol{X}], \boldsymbol{Z}\rangle \neq 0, \quad \forall \boldsymbol{X} \in \mathcal{N}(\mathfrak{L}_{\mathtt{H}}(\lambda_c)). \tag{4.1}$$

To compute (4.1), $\boldsymbol{Z}$ needs to be found as a null solution of $\mathfrak{L}^*_{\mathtt{H}}(\lambda_c)$. The inequality (4.1) is known as the *transversality condition*. Using the expression of adjoint operator (B.4), the dual space restriction (B.3) and inner product (B.1), it turns out that the transversality condition holds; we provide proof in some cases and illustrate graphically in others. The details are provided below.

**4.1.1. Planar critical modes.** As mentioned earlier in §3.3, one of null solutions for critical point $\lambda_k$ corresponding to planar null solutions satisfies the symmetry $\mathtt{Z}_k$. The $\mathtt{Z}_k$ reduced adjoint operator $\mathfrak{L}^*_{Z_k}(\lambda_k)$ is obtained by restricting the spaces $\mathcal{D}$, $\mathcal{C}$ using the condition (3.21) (that is constructing $\mathcal{D}_{\mathtt{Z}_k}, \mathcal{C}_{\mathtt{Z}_k}$), and applying (3.21) to (B.4). Therefore, the $\mathtt{Z}_k$-reduced adjoint operator is

$$\begin{aligned}\mathfrak{L}^*_{\mathtt{Z}_k}(\lambda_k)[\boldsymbol{Z}] = \Big(&-12(1-\nu)\nabla\cdot\mathbf{S} - 12\nu\nabla(\mathbf{S}:\mathbf{I}), 0, -\tfrac{1}{2}\left(\nabla\boldsymbol{u} + \nabla\boldsymbol{u}^\top\right) - \mathbf{S}, \mathbf{0}, \mathbf{0},\\ &2\mathbf{C}\boldsymbol{q}'' - 2\alpha(\gamma-1)(\boldsymbol{t}\otimes\boldsymbol{l} + \boldsymbol{l}\otimes\boldsymbol{t})\boldsymbol{q}' - 2\boldsymbol{t}\wedge\boldsymbol{h}, \boldsymbol{q}\wedge\boldsymbol{t} - \boldsymbol{u}'\Big)^\top,\end{aligned} \tag{4.2}$$

where $\mathbf{C}$ is defined by $(2.1)_2$, $\boldsymbol{Z}$ is defined by (1.7), and $\boldsymbol{q}, \boldsymbol{h}$ have the form

$$\boldsymbol{q} = q_3\boldsymbol{e}_3, \quad \boldsymbol{h} = h_r\boldsymbol{e}_r + h_\theta\boldsymbol{e}_\theta. \tag{4.3}$$

The nontrivial components, defined on $\Omega$ and $\overline{\Omega}$, respectively, of $\mathfrak{L}^*[\boldsymbol{Z}] = \mathbf{0}$ are

$$-12(1-\nu)\nabla\cdot\mathbf{S} - 12\nu\nabla(\mathbf{S}:\mathbf{I}) = \mathbf{0}, \quad -\tfrac{1}{2}\left(\nabla\boldsymbol{u} + \nabla\boldsymbol{u}^\top\right) - \mathbf{S} = \mathbf{0}. \tag{4.4}$$

Substituting $(4.4)_2$ in $(4.4)_1$, and separating polar components, we get

$$\begin{aligned}-2u_r + (1-\nu)u_{r,\theta\theta} + (\nu-3)u_{\theta,\theta} + r(2u_{r,r} + 2ru_{r,rr} + (1+\nu)u_{\theta,\theta r}) &= 0,\\ r((1+\nu)u_{r,\theta r} + (1-\nu)u_{\theta,r} + (1-\nu)ru_{\theta,rr}) + (3-\nu)u_{r,\theta} - (1-\nu)u_\theta + 2u_{\theta,\theta\theta} &= 0,\end{aligned} \tag{4.5}$$

on $\Omega$. The nontrivial (polar) components, defined on $\Gamma$, of $\mathfrak{L}^*[\boldsymbol{Z}] = \mathbf{0}$ are

$$2\alpha^2\beta q_{3,\theta\theta} + 2h_r = 0, \quad -q_3 + \alpha u_\theta - \alpha u_{r,\theta} = 0, \quad u_r + u_{\theta,\theta} = 0. \tag{4.6}$$

Further, $\boldsymbol{Z}$ in (4.2) is itself restricted. On applying (3.21) on (B.3), it is found that only (B.3a) is nontrivial. Substituting (4.3) in (B.3a), and separating the polar components, we get the following two scalar equations on $\Gamma$:

$$12(S_{rr} + \nu S_{\theta\theta}) + 12(1+\nu)(1-\lambda)q_{3,\theta} - \alpha h_{r,\theta} + \alpha h_\theta = 0, \tag{4.7a}$$

$$12(1-\nu)S_{r\theta} + 12(1+\nu)(1-\lambda)q_3 - \alpha h_{\theta,\theta} - \alpha h_r = 0. \tag{4.7b}$$

Substituting $h_r$ and $q_3$ from $(4.6)_1$ and $(4.6)_2$ in (4.7a), and using $(4.4)_2$, gives $h_\theta$ in terms of $u_r$ and $u_\theta$ on $\Gamma$. Substituting $h_r$, $q_3$, and $h_\theta$ in (4.7b) gives a boundary condition for $u_r$ and $u_\theta$. The second boundary condition for $u_r$ and $u_\theta$ is $(4.6)_3$.

The solution for $u_r$ and $u_\theta$, corresponding to $\mathfrak{L}^*_{\mathtt{Z}_k}(\lambda_k)[\boldsymbol{Z}_k] = \mathbf{0}$, can be obtained by separation of variables (Fourier expansion in $\theta$) in domain equations (i.e., equations on $\Omega$) (4.5) and boundary conditions (i.e., equations on $\Gamma$) (4.6) and (4.7). The general solution (of (4.5), with boundary conditions derived from (4.6), (4.7)) is given by

$$\begin{aligned} u_r(r,\theta) &= (-A_k r^{k-1} - C_k \tfrac{k-2+\nu(k+2)}{4+k(1+\nu)} r^{k+1}) \cos k\theta, \\ u_\theta(r,\theta) &= (A_k r^{k-1} + C_k r^{k+1}) \sin k\theta, \end{aligned} \tag{4.8}$$

where $A_k : C_k = -(k+1)(-2\nu + \nu k + k + 2) : \alpha^2(k-1)(\nu k + k + 4)$, on $\overline{\Omega}$.

Rest of the non-zero components of $\boldsymbol{Z}_k$ (with $\mathfrak{L}^*_{\mathtt{Z}_k}(\lambda_k)[\boldsymbol{Z}_k] = \mathbf{0}$), i.e., $\mathbf{S}$, $\boldsymbol{q}$ and $\boldsymbol{h}$ are determined using $(4.4)_2$, $(4.6)_2$, $(4.6)_1$, and (4.7a). The components of $\boldsymbol{Z}_k$ which are zero can be found by inspection of (3.21) due to the structure of $\mathtt{Z}_k$ symmetry-reduced $\mathcal{C}$. All the non-zero components of $\boldsymbol{Z}_k$ are stated in (C.1). It is observed that the expression of $\boldsymbol{Z}_k$ has the same form as that of $\boldsymbol{X}_1$ stated in (A.2) with naturally different $r$ dependent coefficient functions.

Applying the $\mathtt{Z}_k$ symmetry-based reduction condition (3.21) on (2.1), and differentiating with respect to $\lambda$, we get the following expression which occurs in the transversality condition:

$$D_\lambda \mathfrak{L}_{\mathtt{Z}_n}(\lambda_k)[\boldsymbol{X}] = \Big(\mathbf{0}, 0, \mathbf{0}, \mathbf{0}, \mathbf{0}, -\frac{12}{\alpha}(1+\nu)(\lambda-1)\boldsymbol{t} \wedge \boldsymbol{v}', \mathbf{0}\Big)^\top. \tag{4.9}$$

Substituting (4.9) in the left hand side of transversality condition (4.1) along with the expressions for $\boldsymbol{Z}_k$ (C.1), we get

$$\langle \mathrm{D}_\lambda \mathfrak{L}_{\mathtt{Z}_k}(\lambda_k)[\boldsymbol{X}_k], \boldsymbol{Z}_k \rangle = -\sqrt{\langle u_\theta, u_\theta \rangle \langle v_\theta, v_\theta \rangle} \frac{4\pi(k-1)^2(\nu-3)^2\alpha^{3-2k}}{(k(1+\nu)+2(1-\nu))^2}. \tag{4.10}$$

Therefore, the transversality condition (4.1) holds for $k \geq 2$. Hence by Theorem 3.4, all planar critical points $\lambda_k$ are bifurcation points.

**4.1.2. Non-planar critical modes.** As mentioned earlier in §3.3, the non-planar modes have $\mathtt{D}_k$ symmetry, and in some cases even $\mathtt{O}(2)$ symmetry. The operations described in next paragraphs are carried out for $\mathtt{D}_k$ symmetry. The operations for $\mathtt{O}(2)$ symmetry are identical except for the absence of $\theta$ dependent terms.

The $\mathtt{D}_k$ reduced adjoint operator $\mathfrak{L}^*_{\mathtt{D}_k}(\lambda_k)$ is obtained by restricting the spaces $\mathcal{D}$ and $\mathcal{C}$, using the same condition (3.17) (that is constructing $\mathcal{D}_{\mathtt{D}_k}, \mathcal{C}_{\mathtt{D}_k}$). The formal expression of the operator $\mathfrak{L}^*_{\mathtt{D}_k}(\lambda_k)$ is same as that of $\mathfrak{L}^*(\lambda_k)$ (B.4). As $\mathtt{D}_k \subset \mathtt{Z}_k$, and the planar and non-planar critical parameter values come from a decoupled set of equations and the transversality condition for planar null solutions is already evaluated using $\mathtt{Z}_k$-reduced problem, we need to consider only non-planar dual vectors for analyzing the transversality condition.

Out of all the component equations stated in (B.4), the equations for out-of-plane bending and out-of-plane displacement on $\Omega$, corresponding to $\mathfrak{L}^*(\lambda)[\boldsymbol{Z}] = \mathbf{0}$, are

$$\tfrac{h^2}{\lambda^2}(1-\nu)\nabla\cdot(\nabla\cdot\mathbf{T}) + \tfrac{h^2}{\lambda^2}\nu\Delta(\mathbf{T}:\mathbf{I}) + \tfrac{12}{\lambda}(1+\nu)(1-\lambda)\nabla\cdot\boldsymbol{p} = 0, \tag{4.11}$$

$$\nabla^2 w - \mathbf{T} = \mathbf{0}, \quad \boldsymbol{p} - \nabla w = \mathbf{0}. \tag{4.12}$$

Substituting $(4.12)_1$ and $(4.12)_2$ in (4.11), we get, in polar coordinates,

$$\begin{aligned}&-r\left(r\left(r^2w_{,rrrr}+2rw_{,rrr}-w_{,rr}+2w_{,rr\theta\theta}\right)+w_{,r}-2w_{,r\theta\theta}\right)-\\&-4w_{,\theta\theta}-w_{,\theta\theta\theta\theta}+12\tfrac{1+\nu}{h^2}(\lambda-\lambda^2)\left(r^4w_{,rr}+r^3w_{,r}+r^2w_{,\theta\theta}\right)=0,\end{aligned}\tag{4.13}$$

on $\Omega$. The general solution for $w$ (regular as $r\to 0$), by solving (4.13), is given by

$$w(r,\theta)=\begin{cases}(A_kr^k+C_kI_k(\sqrt{|a_\lambda|}r))\cos k\theta, & 0<\lambda<1,\\ (A_kr^k+C_kJ_k(\sqrt{|a_\lambda|}r))\cos k\theta, & \lambda>1,\end{cases}\quad a_\lambda=\frac{12(1+\nu)(1-\lambda)\lambda}{h^2},\tag{4.14}$$

where $\lambda=\lambda_{k,n}$ is the $n$th critical point for out-of-plane null solution of $\mathfrak{L}(\lambda)$ for $\mathtt{D}_k$ symmetric modes ($n=1$ for $\lambda<1$, and $n=1,2,\ldots$ for $\lambda>1$). On substituting $\boldsymbol{T}$ from $(4.12)_1$ in (B.3c) and substituting general solution for $w$, gives $A_k$ in terms of $C_k$.

Out of all the component equations stated in (B.4), for out-of-plane bending and out-of-plane displacement on $\Gamma$, we get

$$2\alpha^2\beta(q_{r,\theta\theta}-q_{\theta,\theta}-q_r)-2\alpha^2(\gamma-\beta)(q_{\theta,\theta}+q_r)-2h_3=0,\tag{4.15a}$$

$$2\gamma\alpha^2(q_{\theta,\theta\theta}+2q_{r,\theta}-q_\theta)-2\alpha^2(\gamma-\beta)(q_{r,\theta}-q_\theta)=0,\quad q_r-\alpha w_{,\theta}=0.\tag{4.15b}$$

Due to above mentioned consideration that only non-planar dual vectors need to be investigated for analyzing the transversality condition, the other non-zero components of $\boldsymbol{Z}_k$ (which represents any of the $\boldsymbol{Z}_{k,n}$s since $n$ does not enter into the expressions), namely $\mathbf{T},\boldsymbol{p},q_r,q_\theta$ and $h_3$, are determined using $(4.12)_1$, $(4.12)_2$, $(4.15b)_2$, $(4.15b)_1$ and (4.15a), respectively. The null solution of solution of $\mathfrak{L}^*_{\mathtt{D}_k}(\lambda_{k,n})$, denoted by $\boldsymbol{Z}_k$ has the same form as $\boldsymbol{X}_1$ of (A.1) but with different radial coefficient functions.

Similar to (4.9), applying the $\mathtt{D}_k$ symmetry reduction condition (3.17) on (2.1), and differentiating with respect to $\lambda$, get

$$\begin{aligned}\mathrm{D}_\lambda\mathfrak{L}_{\mathtt{D}_k}(\lambda)[\boldsymbol{X}]=\Big(&\mathbf{0},0,\mathbf{0},\tfrac{2h^2}{\lambda^3}((1-\nu)\nabla^2z+\nu\,\mathbf{I}\,\Delta z),\tfrac{12}{\lambda^2}(1+\nu)\nabla z,\\&-\tfrac{12}{\alpha}(1+\nu)\boldsymbol{t}\wedge(\boldsymbol{v}'+z'\boldsymbol{e}_3),\mathbf{0}\Big)^\top,\end{aligned}\tag{4.16}$$

where $\lambda=\lambda_{k,n}$ as mentioned alongside (4.14). Substituting (4.16) in left hand side of (4.1) along with the expression for $\boldsymbol{Z}_k$ (same form as $\boldsymbol{X}_1$), similar to (4.10), we get

$$\begin{aligned}\langle\mathrm{D}_\lambda\mathfrak{L}_{\mathtt{D}_k}(\lambda_k)[\boldsymbol{X}_k],\boldsymbol{Z}_k\rangle=\sqrt{\frac{\langle w,w\rangle}{\langle z,z\rangle}}\Big(&\int_\Omega\tfrac{2h^2}{\lambda^3}W(\nabla^2z)\mathrm{d}A+\\&+\int_\Omega\tfrac{12}{\lambda^2}(1+\nu)\nabla z\cdot\nabla z\mathrm{d}A-\int_\Gamma\tfrac{12}{\alpha}(1+\nu)z'^2\mathrm{d}l\Big).\end{aligned}\tag{4.17}$$

where $W(\mathbf{A})=\nu(\mathrm{tr}\mathbf{A})^2+(1-\nu)\mathbf{A}:\mathbf{A}$, for an arbitrary tensor $\mathbf{A}$, $\lambda=\lambda_{k,n}$ (4.14).

In contrast to (4.10), in this case this expression (4.17) is slightly more complicated. The first two terms of (4.17) depend on the deformation of the plate and the third term is dependent on the rod center-line deformation. The change in sign of $\langle\mathrm{D}_\lambda\mathfrak{L}_{\mathtt{D}_k}(\lambda_k)[\boldsymbol{X}_k],\boldsymbol{Z}_k\rangle$ in (4.17), due to the minus sign in the third term, signifies the change in the nature of bifurcation (roughly speaking, a plate dominant buckling mode versus a rod dominant buckling mode) of the system. As the positive definite terms are dependent only on the plate deformation while the third term depends only on the rod, the expression (4.17) is not generally zero; the numerical values are evaluated in a later section. We can only state that the critical points, where (4.17) is non-zero, are certainly bifurcation points.

*Remark* 4.1. The transversality condition holds rigorously for $\mathtt{Z}_k$ symmetry using equation (4.10), as the right hand side in (4.10) is non-zero for all allowable parameter values. For $\mathtt{O}(2)$ symmetric modes, the displacement is constant on boundary, making third term of equation (4.17) as zero, hence the transversality condition also holds rigorously in this case as well.

*Remark* 4.2. For $\mathtt{D}_k$ symmetric modes, for $\lambda > 1$, the transversality condition (4.17) can be satisfied rigorously if the conjecture holds that the null solutions of this type are same as those of circular plate simply supported at its edge; in this case the third term of equation (4.17) is zero, while the first two terms are positive definite implying the transversality condition. However, we do not have a proof of this conjecture but we have found numerically that the critical values and null solutions of both problems coincide.

*Remark* 4.3. For $\mathtt{D}_k$ symmetric modes with $\lambda < 1$, the transversality condition is verified numerically; the numerical evaluation of the right hand side in (4.17) is found to be non-zero but these results are not included in this article. However, in this case, these numerical evaluations lead to a finite value of resulting local curvature of $\lambda$, that is $\ddot{\lambda}(0)$, which is evaluated numerically and the plots are included in supplementary S2.

In light of above evidence, and proof in some cases, stated as Remarks 4.1–4.3, therefore, the conclusions of Theorem 3.4 hold for the reduced problem and, consequently, a local bifurcation analysis can be carried out. In the next section, the Lyapunov–Schmidt method [15] is used for the so called (local) post-buckling analysis.

**4.2. Local bifurcating branch.** In a neighbourhood of $(\mathbf{0}, \lambda_c)$, the bifurcating branch of the reduced problem can be written in form $t \mapsto (\boldsymbol{V}(t), \lambda(t)) \in \mathcal{D}_{\mathtt{H}} \times \mathbb{R}^+$. We have the following expressions, in accordance with notation used in this article, for derivatives of $\lambda(t)$ at the bifurcation point, based on the exposition in [15] (equations I.6.3 and I.6.11):

$$\lambda(0) = \lambda_c, \dot{\lambda}(0) = -\frac{1}{2}\frac{\langle \mathrm{D}^2_{\boldsymbol{V}}\mathfrak{F}_{\mathtt{H}}[\boldsymbol{X}_{\mathtt{H}}, \boldsymbol{X}_{\mathtt{H}}], \boldsymbol{Z}_{\mathtt{H}}\rangle}{\langle \mathrm{D}_\lambda \mathfrak{L}_{\mathtt{H}}(\lambda_c)[\boldsymbol{X}_{\mathtt{H}}], \boldsymbol{Z}_{\mathtt{H}}\rangle}, \ddot{\lambda}(0) = -\frac{1}{3}\frac{\langle \boldsymbol{Z}, \boldsymbol{Z}_{\mathtt{H}}\rangle}{\langle \mathrm{D}_\lambda \mathfrak{L}_{\mathtt{H}}(\lambda_c)[\boldsymbol{X}_{\mathtt{H}}], \boldsymbol{Z}_{\mathtt{H}}\rangle} \quad (4.18a)$$

$$\boldsymbol{Z} := \mathrm{D}^3_{\boldsymbol{V}}\mathfrak{F}_{\mathtt{H}}(\mathbf{0}, \lambda_c)[\boldsymbol{X}_{\mathtt{H}}, \boldsymbol{X}_{\mathtt{H}}, \boldsymbol{X}_{\mathtt{H}}] - 3\mathrm{D}^2_{\boldsymbol{V}}\mathfrak{F}_{\mathtt{H}}(\mathbf{0}, \lambda_c)[\boldsymbol{X}_{\mathtt{H}}, \boldsymbol{Y}], \quad (4.18b)$$

$$\boldsymbol{Y} := (\mathcal{I} - \mathcal{A})(\mathfrak{L}_{\mathtt{H}}(\lambda_c))^{-1}(\mathcal{I} - \mathcal{B})[\mathrm{D}^2_{\boldsymbol{V}}\mathfrak{F}_{\mathtt{H}}(\mathbf{0}, \lambda_c)[\boldsymbol{X}_{\mathtt{H}}, \boldsymbol{X}_{\mathtt{H}}]], \quad (4.18c)$$

where $\boldsymbol{X}_{\mathtt{H}}$ is the null solution of $\mathfrak{L}_{\mathtt{H}}(\lambda_c)$ and $\boldsymbol{Z}_{\mathtt{H}}$ is the null solution of $\mathfrak{L}^*_{\mathtt{H}}(\lambda_c)$. Note

$$\lambda(t) = \lambda(0) + t\dot{\lambda}(0) + \tfrac{t^2}{2}\ddot{\lambda}(0) + o(t^2), \text{ and } \boldsymbol{V}(t) = \mathbf{0} + t\boldsymbol{X}_{\mathtt{H}} + \frac{1}{2}t^2(-\boldsymbol{Y}) + o(t^2), \quad (4.19)$$

as $t \to 0$.

The evaluation of the local bifurcation parameters in (4.18) is carried out for three symmetry subgroups (defined in (3.2a), (3.2b), and (3.2c)) in the following subsections. The insight obtained after these evaluations is that the second order approximation add only an in-plane contribution only, even for out-of-plane buckling. These calculations involve analytical results for planar buckling and semi-analytical results for out-of-plane buckling, due to the presence of transcendental terms in the equation for $\lambda_c$ in case of latter; thus, a numerical method is needed to find the bifurcation points for out-of-plane buckling.

**4.2.1. For planar solutions.** Recall that $\mathfrak{L}_{\mathtt{Z}_k}(\lambda_k)[\boldsymbol{X}_k] = \mathbf{0}$ and $\mathfrak{L}^*_{\mathtt{Z}_k}(\lambda_k)[\boldsymbol{Z}_k] = \mathbf{0}$. Applying (3.21) on (1.8), $\mathtt{Z}_k$ symmetry-reduced nonlinear operator is given by

$$\begin{aligned}\mathfrak{F}_{\mathtt{Z}_k}(\boldsymbol{V},\lambda) = \Big(\nabla\cdot\mathbf{N}, 0, 6\left((1-\nu)\left(\nabla\boldsymbol{v}+\nabla\boldsymbol{v}^\top\right)+2\nu\nabla\cdot\boldsymbol{v}\mathbf{I}\right)-\mathbf{N},\mathbf{0},\mathbf{0},\\ \boldsymbol{m}' - (\boldsymbol{n}+\tfrac{12}{\alpha}(1+\nu)(\lambda-1)\boldsymbol{t})\wedge(\boldsymbol{t}+\boldsymbol{v}'), \boldsymbol{t}+\boldsymbol{v}'-\mathtt{R}(\boldsymbol{g})[\boldsymbol{t}]\Big)^\top,\end{aligned} \tag{4.20}$$

where $\mathtt{R}(\boldsymbol{g})$ is defined in $(1.10c)_2$ and $\boldsymbol{V}\in\mathcal{D}_{\mathtt{Z}_k}, \lambda>0$. Differentiating $\mathfrak{F}_{\mathtt{Z}_k}$ twice with respect to $\boldsymbol{V}$ along $\boldsymbol{X}_k$ (corresponding to the critical value $\lambda_k$, stated as $\boldsymbol{X}_1$ in (A.2)) and evaluating it at the trivial solution $\boldsymbol{V}=\mathbf{0}$, we get the following expression:

$$\mathrm{D}^2_V\mathfrak{F}_{\mathtt{Z}_k}(\mathbf{0},\lambda_k)[\boldsymbol{X}_k,\boldsymbol{X}_k] = \Big(\mathbf{0},0,\mathbf{0},\mathbf{0},\mathbf{0},-2\boldsymbol{n}\wedge\boldsymbol{v}', 4\mathrm{s}_3^2\boldsymbol{t}\Big)^\top. \tag{4.21}$$

After substituting (4.21) and the expression of $\boldsymbol{Z}_k$, from (C.1), in the numerator of $(4.18a)_2$, and then expressing the result in polar coordinates, we get

$$\begin{aligned}\langle \mathrm{D}^2_V\mathfrak{F}_{\mathtt{Z}_k}(\mathbf{0},\lambda_k)[\boldsymbol{X}_k,\boldsymbol{X}_k],\boldsymbol{Z}_k\rangle = &-2\int_0^{2\pi}\left(n_r(v_{\theta,\theta}+v_r)-n_\theta(v_{r,\theta}-v_\theta)\right)q_3\,\mathrm{d}\theta\\ &+4\int_0^{2\pi}\mathrm{s}_3^2h_\theta\,\mathrm{d}\theta.\end{aligned} \tag{4.22}$$

Using the restrictions on $\mathtt{Z}_k$ symmetry-reduced space (3.21) for $\boldsymbol{X}_k$ and $\boldsymbol{Z}_k$, we get

$$\langle \mathrm{D}^2_V\mathfrak{F}_{\mathtt{Z}_k}(\mathbf{0},\lambda_k)[\boldsymbol{X}_k,\boldsymbol{X}_k],\boldsymbol{Z}_k\rangle = 0. \tag{4.23}$$

By substituting a part of the transversality condition, stated as (4.10), and (4.23) in $(4.18a)_2$, we find that $\dot\lambda(0)=0$. The reduced spaces $\mathcal{D}_{\mathtt{Z}_k}$ and $\mathcal{R}_{\mathtt{Z}_k}$ are decomposed as $\mathcal{D}_{\mathtt{Z}_k} = \mathcal{N}(\mathfrak{L}_{\mathtt{Z}_k}(\lambda_k))\oplus\mathcal{D}_{0\mathtt{Z}_k}, \mathcal{R}_{\mathtt{Z}_k} = \mathcal{R}_{0\mathtt{Z}_k}\oplus\mathcal{R}(\mathfrak{L}_{\mathtt{Z}_k}(\lambda_k))$, where due to symmetry reduction $\mathcal{N}(\mathfrak{L}_{\mathtt{Z}_k}(\lambda_k))$ and $\mathcal{R}_{0\mathtt{Z}_k}$ are one dimensional. Projections $\mathcal{A}:\mathcal{D}_{\mathtt{Z}_k}\to\mathcal{N}(\mathfrak{L}_{\mathtt{Z}_k}(\lambda_k))$ and $\mathcal{B}:\mathcal{R}_{\mathtt{Z}_k}\to\mathcal{R}_{0\mathtt{Z}_k}$ onto these one dimensional spaces are defined by

$$\mathcal{A}(\boldsymbol{V}) := \langle\boldsymbol{V},\boldsymbol{X}_k\rangle\boldsymbol{X}_k,\quad \mathcal{B}(\boldsymbol{W}) := \langle\boldsymbol{W},\boldsymbol{Z}_k\rangle\boldsymbol{Z}_k, \tag{4.24}$$

where $\boldsymbol{X}_k$ and $\boldsymbol{Z}_k$ are null solutions of $\mathfrak{L}_{\mathtt{Z}_k}(\lambda_k)$ and $\mathfrak{L}^*_{\mathtt{Z}_k}(\lambda_k)$, such that $\langle\boldsymbol{X}_k,\boldsymbol{X}_k\rangle = 1, \langle\boldsymbol{Z}_k,\boldsymbol{Z}_k\rangle = 1$, respectively.

Differentiating (4.20) thrice with respect to $\boldsymbol{V}$ along $\boldsymbol{X}_k$, and evaluating at trivial solution $\boldsymbol{V}=\mathbf{0}$, gives the following expression:

$$\mathrm{D}^3_V\mathfrak{F}_{\mathtt{Z}_k}(\mathbf{0},\lambda_k)[\boldsymbol{X}_k,\boldsymbol{X}_k,\boldsymbol{X}_k] = \Big(\mathbf{0},0,\mathbf{0},\mathbf{0},\mathbf{0},-12\beta(\mathrm{s}_3^2\mathrm{s}_3''+2\mathrm{s}_3\mathrm{s}_3'^2)\boldsymbol{e}_3,-12\mathrm{s}_3^3\boldsymbol{l}\Big)^\top. \tag{4.25}$$

Using (4.23) and (4.24), we get the following expression for a part of the second term in (4.18b),

$$(\mathcal{I}-\mathcal{B})\mathrm{D}^2_V\mathfrak{F}_{\mathtt{Z}_k}(\mathbf{0},\lambda_k)[\boldsymbol{X}_k,\boldsymbol{X}_k] = \mathrm{D}^2_V\mathfrak{F}_{\mathtt{Z}_k}(\mathbf{0},\lambda_k)[\boldsymbol{X}_k,\boldsymbol{X}_k]. \tag{4.26}$$

For evaluating the second term in (4.18b), the inverse of linear operator $\mathfrak{L}_{\mathtt{Z}_k}(\lambda_k)$ acting on (4.26) needs to be found. Towards that end, we consider the following.

Let $\boldsymbol{Y}\in\mathcal{D}_{\mathtt{Z}_k}$ be such that

$$\mathfrak{L}_{\mathtt{Z}_k}(\lambda_k)[\boldsymbol{Y}] = \mathrm{D}^2_V\mathfrak{F}_{\mathtt{Z}_k}(\mathbf{0},\lambda_k)[\boldsymbol{X}_k,\boldsymbol{X}_k]. \tag{4.27}$$

Due to the structure of $\mathcal{D}_{\mathtt{Z}_k}$, we have the form (similar to the notation (3.9))

$$\boldsymbol{Y} = \Big(\boldsymbol{y}^{(1)},0,\mathbf{Y}^{(3)},\mathbf{0},\mathbf{0},\boldsymbol{y}^{(6)},\boldsymbol{y}^{(7)}\Big)^\top. \tag{4.28}$$

On substituting such $\boldsymbol{Y}$ (4.28) in (2.1), and (4.21) in (4.27), the following set of non-trivial equations are obtained (other equations are of the form $0 = 0$):

$$\nabla \cdot \mathbf{Y}^{(3)} = \mathbf{0}, \quad 6((1-\nu)(\nabla \boldsymbol{y}^{(1)} + \nabla \boldsymbol{y}^{(1)\top}) + 2\nu \nabla \cdot \boldsymbol{y}^{(1)} \mathbf{I}) - \mathbf{Y}^{(3)} = \mathbf{0}, \tag{4.29a}$$

on $\Omega$, and

$$2\mathbf{C}[\boldsymbol{y}^{(6)''}] - 2\alpha(\gamma - 1)(\boldsymbol{t} \otimes \boldsymbol{l} + \boldsymbol{l} \otimes \boldsymbol{t})[\boldsymbol{y}^{(6)'}] + \\ + \boldsymbol{t} \wedge (\boldsymbol{y}^{(7)} - \tfrac{12}{\alpha}(1+\nu)(\lambda - 1)\boldsymbol{y}^{(1)'}) = -2\boldsymbol{n} \wedge \boldsymbol{v}', \tag{4.30a}$$

$$\boldsymbol{y}^{(1)'} - 2\boldsymbol{y}^{(6)} \wedge \boldsymbol{t} = 4\mathrm{s}_3^2 \boldsymbol{t}, \tag{4.30b}$$

on $\partial\Omega$, where $\mathbf{C}$ is defined by $(2.1)_2$.

Applying restriction (3.21) on first condition in definition of $\mathcal{D}$ (other two conditions are trivially satisfied), the boundary condition on $\boldsymbol{Y}$ (4.28) is found as

$$\boldsymbol{y}^{(7)\prime} - \mathbf{Y}^{(3)} \boldsymbol{l} = \mathbf{0} \text{ on } \Gamma. \tag{4.31}$$

Substituting $\mathbf{Y}^{(3)}$ from $(4.29\text{a})_2$ in $(4.29\text{a})_1$ and separating the polar components (following the scheme explained in Remark 3.3) gives

$$-2y_r^{(1)} + (1-\nu)y_{r,\theta\theta}^{(1)} + (\nu - 3)y_{\theta,\theta}^{(1)} + r(2y_{r,r}^{(1)} + 2ry_{r,rr}^{(1)} + (1+\nu)y_{\theta,\theta r}^{(1)}) = 0, \tag{4.32a}$$

$$(3-\nu)y_{r,\theta}^{(1)} - (1-\nu)y_\theta^{(1)} + r((1+\nu)y_{r,\theta r}^{(1)} + (1-\nu)y_{\theta,r}^{(1)} + \\ + (1-\nu)ry_{\theta,rr}^{(1)}) + 2y_{\theta,\theta\theta}^{(1)} = 0. \tag{4.32b}$$

In polar components, the equations (4.30) and (4.31) on $\Gamma$ can be written as

$$2\alpha^2 \beta y_{3,\theta\theta}^{(6)} + 12(1+\nu)(\lambda - 1)(y_{r,\theta}^{(1)} - y_\theta^{(1)}) - y_r^{(7)} = - \\ -2\alpha(n_r(v_{\theta,\theta} + v_r) - n_\theta(v_{r,\theta} - v_\theta)) \quad , \tag{4.33a}$$

$$\alpha y_{r,\theta}^{(1)} - \alpha y_\theta^{(1)} + 2y_3^{(6)} = 0, \tag{4.33b}$$

$$\alpha y_{\theta,\theta}^{(1)} + \alpha y_r^{(1)} = 4\mathrm{s}_3^2, \tag{4.33c}$$

$$\alpha y_{r,\theta}^{(7)} - \alpha y_\theta^{(7)} - 12\nu\alpha(y_r^{(1)} + y_{\theta,\theta}^{(1)}) - 12y_{r,r}^{(1)} = 0, \tag{4.33d}$$

$$\alpha y_{\theta,\theta}^{(7)} + \alpha y_r^{(7)} - 6(1-\nu)(y_{\theta,r}^{(1)} + \alpha y_{r,\theta}^{(1)} - \alpha y_\theta^{(1)}) = 0. \tag{4.33e}$$

The general solution of (4.32) has infinite number of linear independent solutions. Considering the non-homogeneous parts of boundary condition (4.33c) which has only second order terms of $\boldsymbol{X}_k$ (corresponding to $\lambda_k$, stated as $\boldsymbol{X}_1$ in (A.2) with the form $\Big((v_r, v_\theta), 0, [N_{rr}, N_{\theta\theta}, N_{r\theta}], [(0,0,0)], (0,0), (0,0,s_3), (n_r, n_\theta, 0)\Big)^\top$, the only form of the non-trivial solution is

$$y_r^{(1)}(r,\theta) = Cr + (Ar^{2k+1} + Br^{2k-1})\cos 2k\theta, \tag{4.34a}$$

$$y_\theta^{(1)}(r,\theta) = (-\frac{2 + (1+\nu)k}{(1+\nu)k - (1-\nu)} Ar^{2k+1} - Br^{2k-1})\sin 2k\theta, \tag{4.34b}$$

where $A$, $B$ and $C$ are arbitrary constants. From (4.33b), and subsequently from (4.33a) and (4.33d), using $(4.29\text{a})_2$, we get

$$y_3^{(6)} = \frac{\alpha}{2}(y_\theta^{(1)} - y_{r,\theta}^{(1)}), \quad y_r^{(7)} = 2\alpha^2\beta y_{3,\theta\theta}^{(6)} + 12(1+\nu)(\lambda - 1)(y_{r,\theta}^{(1)} - y_\theta^{(1)}) + \\ + 2\alpha(n_r(v_{\theta,\theta} + v_r) - n_\theta(v_{r,\theta} - v_\theta)), \tag{4.35a}$$

$$y_\theta^{(7)} = y_{r,\theta}^{(7)} - 12\nu(y_r^{(1)} + y_{\theta,\theta}^{(1)}) - \frac{12}{\alpha} y_{r,r}^{(1)}. \tag{4.35b}$$

Expressions (4.34) and (4.35) are substituted in (4.33c) and (4.33e) and coefficients of $\cos 2k\theta$, $\sin 2k\theta$, and $\theta$ independent terms, are collected to obtain a set of three linear equations in $A \equiv A(k)$, $B \equiv B(k)$ and $C \equiv C(k)$. The set of equations are solved to determine the arbitrary constants $A$, $B$ and $C$, which are back substituted in (4.34), (4.35) and $(4.29\text{a})_2$ to obtain $\boldsymbol{Y}$, with the form $\Big(\big(y_r^{(1)}, y_\theta^{(1)}\big), y^{(2)}, \big[Y_{rr}^{(3)}, Y_{\theta\theta}^{(3)}, Y_{r\theta}^{(3)}\big], \big[Y_{rr}^{(4)}, Y_{\theta\theta}^{(4)}, Y_{r\theta}^{(4)}\big], \big(y_r^{(5)}, y_\theta^{(5)}\big), \big(y_r^{(6)}, y_\theta^{(6)}, y_3^{(6)}\big), \big(y_r^{(7)}, y_\theta^{(7)}, y_3^{(7)}\big)\Big)^\top$:

$$\begin{aligned}\boldsymbol{Y} = \Big(&\big(\zeta_1(r) + \zeta_2(r)\cos 2k\theta, \zeta_3(r)\sin 2k\theta\big), 0, \big[\zeta_4(r) + \zeta_5(r)\cos 2k\theta, \zeta_6(r)+\\ &+ \zeta_7(r)\cos 2k\theta, \zeta_8(r)\sin 2k\theta\big], \big[(0,0,0)\big], (0,0), \big(0,0,\zeta_9 \sin 2k\theta\big),\\ &\big(\zeta_{10}\sin 2k\theta, \zeta_{11} + \zeta_{12}\cos 2k\theta, 0\big)\Big)^\top.\end{aligned} \tag{4.36}$$

The expressions of functions (which are constants in some cases) $\{\zeta_i\}_{i=1}^{12}$ have been found explicitly; however, due their cumbersome expressions we do not include them in this main text.

For the other infinite number of terms in the general solution of (4.32), above procedure yields zero particular solution modulo the homogeneous solution $\boldsymbol{X}_k$.

As $\boldsymbol{X}_k$ has the form $\boldsymbol{X}_1$ of (A.2), using projection $\mathcal{A}$ from (4.24), $(\mathcal{I}-\mathcal{A})\boldsymbol{Y} = \boldsymbol{Y}$, where $\boldsymbol{Y}$ is given by (4.36). Using this identity, and using (4.26) and (4.27), we get the second term in (4.18b) (modulo multiplication by $-3$) as

$$\begin{aligned}&\mathrm{D}_V^2\mathfrak{F}_{\mathrm{Z}_k}(\mathbf{0}, \lambda_k)[\boldsymbol{X}_k, (\mathcal{I}-\mathcal{A})(\mathfrak{L}_{\mathrm{Z}_k}(\lambda_k))^{-1}(\mathcal{I}-\mathcal{B})\mathrm{D}_V^2\mathfrak{F}_{\mathrm{Z}_k}(\mathbf{0}, \lambda_k)[\boldsymbol{X}_k, \boldsymbol{X}_k]]\\ &= \mathrm{D}_V^2\mathfrak{F}_{\mathrm{Z}_k}(\mathbf{0}, \lambda_k)[\boldsymbol{X}_k, \boldsymbol{Y}].\end{aligned} \tag{4.37}$$

Evaluating at $\boldsymbol{V} = \mathbf{0}$ the second order (mixed direction) derivative of (4.20) with respect to $\boldsymbol{V}$ along $\boldsymbol{X}_k$ and $\boldsymbol{Y}$, we get second term in (4.18b) as $-3\Big(\mathbf{0}, 0, \mathbf{0}, \mathbf{0}, \mathbf{0}, -\boldsymbol{y}^{(7)} \wedge \boldsymbol{v}' - \boldsymbol{n} \wedge \boldsymbol{y}^{(1)\prime}, 4\mathrm{s}_3 y_3^{(6)} \boldsymbol{t}\Big)^\top$, where the expressions (4.36) need to be substituted further. Finally, this expression, along with the expressions stated in (4.25) and (4.10), are substituted in $(4.18\text{a})_3$ to evaluate $\ddot{\lambda}(0)$. Thus, by (4.19), we obtain the bifurcating branch up to second order (in the context of the nonlinear problem (1.11)) as

$$\boldsymbol{V}(t) = t\boldsymbol{X}_k - \tfrac{1}{2}t^2\boldsymbol{Y} + \boldsymbol{o}(t^2), \lambda(t) = \lambda_k + \tfrac{1}{2}t^2\ddot{\lambda}(0) + o(t^2), \text{ as } t \to 0, \tag{4.38}$$

where $\boldsymbol{Y}$ is given by (4.36) and, corresponding to the critical value $\lambda_k$, $\boldsymbol{X}_k$ is same as that stated as $\boldsymbol{X}_1$ in (A.2). Note that $\boldsymbol{X}_k$ is normalized so that there is no arbitrary scalar in (4.38).

The analysis presented so far in this subsection is rigorous in view of Remarks 4.1.

**4.2.2. For non-planar solutions.** Corresponding to the critical value $\lambda$, recall that $\mathfrak{L}_{\mathrm{D}_k}(\lambda)[\boldsymbol{X}_k] = \mathbf{0}$ and $\mathfrak{L}^*_{\mathrm{D}_k}(\lambda)[\boldsymbol{Z}_k] = \mathbf{0}$, where $\lambda = \lambda_{k,n}$ is the $n$th critical point for out-of-plane null solution of $\mathfrak{L}(\lambda)$ for $\mathrm{D}_k$ symmetric modes ($n = 1$, and it is suppressed in writing, for $\lambda < 1$, and $n = 1, 2, \ldots$ for $\lambda > 1$). The null solution with $\mathrm{D}_k$ symmetry, $\boldsymbol{X}_k$, same as that stated as $\boldsymbol{X}_1$ in (A.1), and $\boldsymbol{Z}_k$, with a structure of the same form as $\boldsymbol{X}_k$, have the following form:

$$\boldsymbol{X}_k = \Big(\mathbf{0}, z, \mathbf{0}, \mathbf{M}, \boldsymbol{f}, \mathrm{s}_r\boldsymbol{e}_r + \mathrm{s}_\theta\boldsymbol{e}_\theta, n_3\boldsymbol{e}_3\Big)^\top, \boldsymbol{Z}_k = \Big(\mathbf{0}, w, \mathbf{0}, \mathbf{T}, \boldsymbol{p}, q_r\boldsymbol{e}_r + q_\theta\boldsymbol{e}_\theta, h_3\boldsymbol{e}_3\Big)^\top. \tag{4.39}$$

Differentiating (1.8) twice with respect to $\boldsymbol{V}$ along $\boldsymbol{X}_k$, and evaluating at the trivial solution $\boldsymbol{V}=\mathbf{0}$, we get the following expression:

$$\begin{aligned}\mathrm{D}^2_{\boldsymbol{V}}\mathfrak{F}_{\mathtt{D}_k}(\mathbf{0},\lambda)[\boldsymbol{X}_k,\boldsymbol{X}_k] = \Big(&\mathbf{0},0,\tfrac{12}{\lambda}(1-\nu)\nabla z\otimes\nabla z+\tfrac{12}{\lambda}\nu\nabla z\cdot\nabla z\mathbf{I},\mathbf{0},\mathbf{0},\\ &((\beta\mathbf{A}+\gamma\mathbf{B})\mathbf{s}')',-4(\boldsymbol{t}\wedge\mathbf{s})\wedge\mathbf{s}\Big)^{\top},\end{aligned} \tag{4.40}$$

where the polar components of tensors $\mathbf{A}$ and $\mathbf{B}$, using (4.39), can be expressed as

$$[\mathbf{A}]^{\mathrm{cyl}} = 4\begin{bmatrix}0 & 0 & \mathrm{s}_\theta\\ 0 & 0 & -2\mathrm{s}_r\\ -\mathrm{s}_\theta & -\mathrm{s}_r & 0\end{bmatrix},\quad [\mathbf{B}]^{\mathrm{cyl}} = 4\begin{bmatrix}0 & 0 & 0\\ 0 & 0 & \mathrm{s}_r\\ 0 & 2\mathrm{s}_r & 0\end{bmatrix}. \tag{4.41}$$

The notation $[\mathbf{A}]^{\mathrm{cyl}}$ stands for the matrix of components of tensor $\mathbf{A}$ in cylindrical (polar) coordinates, i.e., similar to (3.13), $\mathbf{A}=A_{ij}\boldsymbol{e}_i\otimes\boldsymbol{e}_j, i,j\in\{r,\theta,3\}$ using $\boldsymbol{e}_r,\boldsymbol{e}_\theta,\boldsymbol{e}_3$ as standard basis cylindrical basis for $\mathbb{R}^3$.

Substituting (4.40) and (4.39) in the numerator of $(4.18\mathrm{a})_2$, expressing in polar co-ordinates and using the condition (3.17), we get

$$\langle\mathrm{D}^2_{\boldsymbol{V}}\mathfrak{F}_{\mathtt{D}_k}(\mathbf{0},\lambda)[\boldsymbol{X}_k,\boldsymbol{X}_k],\boldsymbol{Z}_k\rangle = 0. \tag{4.42}$$

After substituting (4.42) and (4.17) in $(4.18\mathrm{a})_2$, we find that $\dot{\lambda}(0)=0$, as also concluded in § 4.1.1. Also, similar to § 4.1.1, reduced spaces $\mathcal{D}_{\mathtt{D}_k}$ and $\mathcal{R}_{\mathtt{D}_k}$ are decomposed as $\mathcal{D}_{\mathtt{D}_k}=\mathcal{N}(\mathfrak{L}_{\mathtt{D}_k}(\lambda))\oplus\mathcal{D}_{0\mathtt{D}_k}, \mathcal{R}_{\mathtt{D}_k}=\mathcal{R}_{0\mathtt{D}_k}\oplus\mathcal{R}(\mathfrak{L}_{\mathtt{D}_k}(\lambda))$, where, due to the reduced symmetry, $\mathcal{N}(\mathfrak{L}_{\mathtt{D}_k}(\lambda))$ and $\mathcal{R}_{0\mathtt{D}_k}$ are one dimensional. Projections $\mathcal{A}:\mathcal{D}_{\mathtt{D}_k}\to\mathcal{N}(\mathfrak{L}_{\mathtt{D}_k}(\lambda))$ and $\mathcal{B}:\mathcal{R}_{\mathtt{D}_k}\to\mathcal{R}_{0\mathtt{Z}_k}$ onto these one dimensional spaces are defined in same way as in (4.24), where null solutions are again scaled such that $\langle\boldsymbol{X}_k,\boldsymbol{X}_k\rangle=1, \langle\boldsymbol{Z}_k,\boldsymbol{Z}_k\rangle=1$.

Differentiating (1.8) thrice with respect to $\boldsymbol{V}$ along $\boldsymbol{X}_k$ and evaluating at the trivial solution $\boldsymbol{V}=\mathbf{0}$, we get the following expression:

$$\mathrm{D}^3_{\boldsymbol{V}}\mathfrak{F}_{\mathtt{D}_k}(\mathbf{0},\lambda)[\boldsymbol{X}_k,\boldsymbol{X}_k,\boldsymbol{X}_k] = \Big(\mathbf{0},0,\mathbf{0},\mathbf{0},\mathbf{0},((\beta\bar{\mathbf{A}}+\gamma\bar{\mathbf{B}})\mathbf{s}')',-12(\mathbf{s}\cdot\mathbf{s})(\boldsymbol{t}\wedge\mathbf{s})\Big)^{\top}, \tag{4.43}$$

where the polar components of $\bar{\mathbf{A}}$ and $\bar{\mathbf{B}}$ can be expressed in matrix form as

$$[\bar{\mathbf{A}}]^{\mathrm{cyl}} = 12\begin{bmatrix}-\mathrm{s}_r^2-\mathrm{s}_\theta^2 & -2\mathrm{s}_r\mathrm{s}_\theta & 0\\ 0 & 2\mathrm{s}_r^2 & 0\\ 0 & 0 & -3\mathrm{s}_r^2-\mathrm{s}_\theta^2\end{bmatrix}, [\bar{\mathbf{B}}]^{\mathrm{cyl}} = 12\begin{bmatrix}0 & 2\mathrm{s}_r\mathrm{s}_\theta & 0\\ 0 & -3\mathrm{s}_r^2-\mathrm{s}_\theta^2 & 0\\ 0 & 0 & 2\mathrm{s}_r^2\end{bmatrix}. \tag{4.44}$$

Using (4.42), and the analogue of (4.24), we have the following relation for a part of the second term in (4.18b):

$$(\mathcal{I}-\mathcal{B})\mathrm{D}^2_V\mathfrak{F}_{\mathtt{D}_k}(\mathbf{0},\lambda)[\boldsymbol{X}_k,\boldsymbol{X}_k] = \mathrm{D}^2_V\mathfrak{F}_{\mathtt{D}_k}(\mathbf{0},\lambda)[\boldsymbol{X}_k,\boldsymbol{X}_k]. \tag{4.45}$$

Similar to (4.27), the following is considered:

$$\mathcal{L}_{\mathtt{D}_k}(\lambda)[\boldsymbol{Y}] = \mathrm{D}^2_V\mathfrak{F}_{\mathtt{D}_k}(\mathbf{0},\lambda)[\boldsymbol{X}_k,\boldsymbol{X}_k]. \tag{4.46}$$

Expanding (4.46), we get the following set of equations on $\Omega$:

$$\nabla\cdot\mathbf{Y}^{(3)}=\mathbf{0},\quad \nabla\cdot(\boldsymbol{y}^{(5)}+\nabla\cdot\mathbf{Y}^{(4)})=0, \tag{4.47a}$$

$$6\left((1-\nu)(\nabla\boldsymbol{y}^{(1)}+\nabla\boldsymbol{y}^{(1)\top})+2\nu\nabla\cdot\boldsymbol{y}^{(1)}\mathbf{I}\right)-\mathbf{Y}^{(3)}= \frac{12}{\lambda}(1-\nu)\nabla z\otimes\nabla z+\frac{12}{\lambda}\nu\nabla z\cdot\nabla z\mathbf{I}, \tag{4.47b}$$

$$-\frac{h^2}{\lambda^2}((1-\nu)\nabla^2 y^{(2)}+\nu\Delta y^{(2)}\mathbf{I})-\mathbf{Y}^{(4)}=\mathbf{0}, \tag{4.47c}$$

$$\boldsymbol{y}^{(5)}-\frac{12}{\lambda}(1+\nu)(1-\lambda)\nabla y^{(2)}=\mathbf{0}, \tag{4.47d}$$

and the following set of boundary equations on $\Gamma$:

$$2\mathbf{C}[\boldsymbol{y}^{(6)\prime\prime}]-2\alpha(\gamma-1)(\boldsymbol{t}\otimes\boldsymbol{l}+\boldsymbol{l}\otimes\boldsymbol{t})[\boldsymbol{y}^{(6)\prime}]\quad + +\boldsymbol{t}\wedge(\boldsymbol{y}^{(7)}-\tfrac{12}{\alpha}(1+\nu)(\lambda-1)(\boldsymbol{y}^{(1)\prime}+y^{(2)\prime}\boldsymbol{e}_3))=((\beta\mathbf{A}+\gamma\mathbf{B})\mathrm{s}')', \tag{4.48a}$$

$$\boldsymbol{y}^{(1)\prime}+y^{(2)\prime}\boldsymbol{e}_3-2\boldsymbol{y}^{(6)}\wedge\boldsymbol{t}=-4(\boldsymbol{t}\wedge\mathrm{s})\wedge\mathrm{s}, \tag{4.48b}$$

where the tensor $\mathbf{C}$ is defined by $(2.1)_2$. Further, we restate the conditions in the definition of $\mathcal{D}$ for $\boldsymbol{Y}\in\mathcal{D}_{\mathrm{D}_k}\subset\mathcal{D}$,

$$(\mathtt{I}-\boldsymbol{e}_3\otimes\boldsymbol{e}_3)\boldsymbol{y}^{(7)\prime}-\mathbf{Y}^{(3)}\boldsymbol{l}=\mathbf{0}, \tag{4.49a}$$

$$\boldsymbol{y}^{(7)\prime}\cdot\boldsymbol{e}_3-(\boldsymbol{y}^{(5)}\cdot\boldsymbol{l}+(\nabla\cdot\mathbf{Y}^{(4)})\cdot\boldsymbol{l}+\nabla(Y^{(4)}_{r\theta})\cdot\boldsymbol{t})=0, \tag{4.49b}$$

$$Y^{(4)}_{rr}=0. \tag{4.49c}$$

Substituting $\mathbf{Y}^{(3)}$ from (4.47b) in $(4.47\text{a})_1$, separating into polar components (following the scheme explained in Remark 3.3), we get

$$-2y_r^{(1)}+(1-\nu)y^{(1)}_{r,\theta\theta}+(\nu-3)y^{(1)}_{\theta,\theta}+r(2y^{(1)}_{r,r}+2ry^{(1)}_{r,rr}+(1+\nu)y^{(1)}_{\theta,\theta r})= \frac{2}{r\lambda}\left((1+\nu)(rz_{,r\theta}z_{,\theta}-z^2_{,\theta})+(1-\nu)(rz_{,r}z_{,\theta\theta}+r^2z^2_{,r})+2r^3z_{,r}z_{,rr}\right), \tag{4.50a}$$

$$(3-\nu)y^{(1)}_{r,\theta}-(1-\nu)y^{(1)}_{\theta}+ +r((1+\nu)y^{(1)}_{r,\theta r}+(1-\nu)y^{(1)}_{\theta,r}+(1-\nu)ry^{(1)}_{\theta,rr})+2y^{(1)}_{\theta,\theta\theta}= \frac{2}{r\lambda}\left(2z_{,\theta}z_{,\theta\theta}+(1-\nu)(rz_{,\theta}z_{,r}+r^2z_{,\theta}z_{,rr})+(1+\nu)r^2z_{,r}z_{,r\theta}\right). \tag{4.50b}$$

Substituting $\mathbf{Y}^{(4)}$ and $\boldsymbol{y}^{(5)}$ from (4.47c) and (4.47d), respectively, in $(4.47\text{a})_2$ gives the following equation in $y^{(2)}$:

$$r\left(r(r^2y^{(2)}_{,rrrr}+2ry^{(2)}_{,rrr}-y^{(2)}_{,rr}+2y^{(2)}_{,rr\theta\theta})+y^{(2)}_{,r}-2y^{(2)}_{,r\theta\theta}\right)+ +4y^{(2)}_{,\theta\theta}+y^{(2)}_{,\theta\theta\theta\theta}-12\tfrac{1+\nu}{h^2}(\lambda-\lambda^2)(r^4y^{(2)}_{,rr}+r^3y^{(2)}_{,r}+r^2y^{(2)}_{,\theta\theta})=0. \tag{4.51}$$

Separating (4.48) into polar components, we get

$$-2\alpha^2(-\beta y^{(6)}_{r,\theta\theta}+\gamma y^{(6)}_r+(\beta+\gamma)y^{(6)}_{\theta,\theta})+y_3^{(7)}-12(\lambda-1)(\nu+1)y^{(2)}_{,\theta}=0, \tag{4.52}$$

$$2\alpha^2((1+\gamma)y^{(6)}_{r,\theta}+\gamma y^{(6)}_{\theta,\theta\theta}-\beta y^{(6)}_\theta)=0, \tag{4.53}$$

$$12(\lambda-1)(\nu+1)(y^{(1)}_{r,\theta}-y^{(1)}_\theta)+2\alpha^2\beta y^{(6)}_{3,\theta\theta}-y^{(7)}_r=-4\alpha^2\times \times(\beta\mathrm{s}_\theta\mathrm{s}_{r,\theta\theta}+(\beta-2\gamma)\mathrm{s}_r(2\mathrm{s}_{r,\theta}+\mathrm{s}_{\theta,\theta\theta})-2\mathrm{s}_{\theta,\theta}((\gamma-\beta)\mathrm{s}_{r,\theta}+\mathrm{s}_\theta)), \tag{4.54}$$

$$\alpha(y^{(1)}_{r,\theta}-y^{(1)}_\theta)+2y_3^{(6)}=-4\mathrm{s}_r\mathrm{s}_\theta, \tag{4.55}$$

$$\alpha(y^{(1)}_r+y^{(1)}_{\theta,\theta})=4\mathrm{s}_r^2, \tag{4.56}$$

$$\alpha y^{(2)}_{,\theta}-2y^{(6)}_r=0. \tag{4.57}$$

Substituting $\mathbf{Y}^{(3)}$, $\mathbf{Y}^{(4)}$ and $\boldsymbol{y}^{(5)}$ from (4.47b), (4.47c) and (4.47d), respectively in (4.49) and separating into polar components (following Remark 3.3) gives

$$-\alpha(12\nu(y_r^{(1)}+y_{\theta,\theta}^{(1)})-y_{r,\theta}^{(7)}+y_\theta^{(7)})-12y_{r,r}^{(1)}=-\tfrac{12}{\lambda}(\alpha^2\nu z_{,\theta}^2+z_{,r}^2), \quad (4.58a)$$

$$-6(1-\nu)(\alpha y_{r,\theta}^{(1)}+y_{\theta,r}^{(1)}-\alpha y_\theta^{(1)})+\alpha(y_r^{(7)}+y_{\theta,\theta}^{(7)})=-\tfrac{12}{\lambda}\alpha(1-\nu)z_{,\theta}z_{,r}, \quad (4.58b)$$

$$-(3-\nu)\tfrac{h^2\alpha^3}{\lambda^2}y_{,\theta\theta}^{(2)}+(2-\nu)\tfrac{h^2\alpha^2}{\lambda^2}y_{,r\theta\theta}^{(2)}+\tfrac{h^2}{\lambda^2}y_{,rrr}^{(2)}+$$
$$+(\tfrac{12(1+\nu)(\lambda-1)}{\lambda}-\tfrac{h^2\alpha^2}{\lambda^2})y_{,r}^{(2)}+\tfrac{h^2\alpha}{\lambda^2}y_{,rr}^{(2)}+\alpha y_{3,\theta}^{(7)}=0, \quad (4.58c)$$

$$\nu(\alpha^2 y_{,\theta\theta}^{(2)}+\alpha y_{,r}^{(2)})+y_{,rr}^{(2)}=0. \quad (4.58d)$$

It can be seen that the non-homogeneous terms only appear in planar equations which are decoupled from non-planar equations. Only zero solutions satisfy the homogeneous equations as we are looking for solutions which are complementary to the null space. The set of non-homogeneous equations (4.50) have the following general solution (including only those which are regular as $r\to 0$):

$$y_r^{(1)}(r,\theta)=Cr+\phi_1(r)+\big(Ar^{2k+1}+Br^{2k-1}+\phi_2(r)\big)\cos 2k\theta, \quad (4.59a)$$

$$y_\theta^{(1)}(r,\theta)=(-\tfrac{2+(1+\nu)k}{(1+\nu)k-(1-\nu)}Ar^{2k+1}-Br^{2k-1}+\psi(r))\sin 2k\theta, \quad (4.59b)$$

where $\phi_1,\phi_2,\psi$ need to be found. Using (4.55),(4.57), and substituting (4.53), we get

$$y_3^{(6)}=-2\mathrm{s}_r\mathrm{s}_\theta-\tfrac{\alpha}{2}(y_{r,\theta}^{(1)}-y_\theta^{(1)}),\quad y_\theta^{(6)}=-(\gamma\tfrac{d^2}{d\theta^2}-\beta)^{-1}\tfrac{(1+\gamma)\alpha}{2}y_{,\theta\theta}^{(2)}. \quad (4.60)$$

From (4.54) and (4.58a), we get

$$y_r^{(7)}=12(\lambda-1)(\nu+1)(y_{r,\theta}^{(1)}-y_\theta^{(1)})+2\alpha^2\beta y_{3,\theta\theta}^{(6)}+4\alpha^2\times$$
$$\times(\beta\mathrm{s}_\theta\mathrm{s}_{r,\theta\theta}+(\beta-2\gamma)\mathrm{s}_r\,(2\mathrm{s}_{r,\theta}+\mathrm{s}_{\theta,\theta\theta})-2\mathrm{s}_{\theta,\theta}((\gamma-\beta)\mathrm{s}_{r,\theta}+\mathrm{s}_\theta)), \quad (4.61a)$$

$$y_\theta^{(7)}=\tfrac{12}{\lambda\alpha}(\alpha^2\nu z_{,\theta}^2+z_{,r}^2)-\tfrac{12}{\alpha}y_{r,r}^{(1)}-(12\nu(y_r^{(1)}+y_{\theta,\theta}^{(1)})-y_{r,\theta}^{(7)}). \quad (4.61b)$$

Expressions (4.59a), (4.59b) are substituted in (4.50). and the coefficients of $\sin 2k\theta$, $\cos 2k\theta$ and $\theta$ independent terms are collected, to obtain the linear ordinary differential equations for $\phi_1,\phi_2,\psi$ (non-homogeneous, with constant coefficients). These equations have been solved explicitly for their particular solution in terms of components of $\boldsymbol{X}_k$; however, again, due their cumbersome expressions we do not include them in the main text of the present article.

Expressions (4.59a), (4.59b), (4.61a) and (4.61b) are substituted in (4.56) and (4.58b), and the coefficients of $\sin 2k\theta$, $\cos 2k\theta$ and $\theta$ independent terms are collected to obtain a set of three linear equations in $A$, $B$ and $C$. The linear equations are solved to obtain the arbitrary constants $A,B,C$.

After executing the procedure described in above two paragraphs (using symbolic computation), all the components of $\boldsymbol{Y}$ can be obtained. It is found that $\boldsymbol{Y}$ has the same form as in (4.36) but with different radial function expressions and different coefficients for functions defined on the boundary. The similarity in the form with (4.36) happens because the non-homogeneous terms of (4.46) only appear corresponding to the planar components of $\boldsymbol{Y}$.

As $\boldsymbol{X}_k$ has the same form as of $\boldsymbol{X}_1$ in (A.1) and $\boldsymbol{Y}$ has the form as in (4.36), using the projection $\mathcal{A}$, analoguous to (4.24), we get $(\mathcal{I}-\mathcal{A})\boldsymbol{Y}=\boldsymbol{Y}$. Therefore, similar to the planar case, we obtain the following expression, corresponding to (4.37), for

non-planar solutions:

$$\mathrm{D}_V^2\mathfrak{F}_{\mathtt{D}_k}(\mathbf{0},\lambda)[\boldsymbol{X}_k,\boldsymbol{Y}] = \Big(\mathbf{0},0,\mathbf{0},\mathbf{0},\mathbf{0},(\tfrac{1}{2}(\beta\mathbf{A}+\gamma\mathbf{B})\boldsymbol{y}^{(6)'}+(\beta\mathbf{A}_y+\gamma\mathbf{B}_y)\mathbf{s}')' \\ -n_3\boldsymbol{e}_3\wedge\boldsymbol{y}^{(1)'}+z'\boldsymbol{e}_3\wedge\boldsymbol{y}^{(7)},-2(\boldsymbol{t}\wedge\boldsymbol{y}^{(6)})\wedge\mathbf{s}-2(\boldsymbol{t}\wedge\mathbf{s})\wedge\boldsymbol{y}^{(6)}\Big)^\top, \tag{4.62}$$

where $\mathbf{A}$ and $\mathbf{B}$ are same as those given in (4.41), and $\mathbf{A}_y$ and $\mathbf{B}_y$ are

$$[\mathbf{A}_y]^{\text{cyl}} = 2\begin{bmatrix} 0 & y_3^{(6)} & 0 \\ 2y_3^{(6)} & 0 & 0 \\ 0 & 0 & 0 \end{bmatrix}, \quad [\mathbf{B}_y]^{\text{cyl}} = 2\begin{bmatrix} 0 & -2y_3^{(6)} & 0 \\ -y_3^{(6)} & 0 & 0 \\ 0 & 0 & 0 \end{bmatrix}, \tag{4.63}$$

using the standard cylindrical basis as that employed in (4.41).

Above procedure leads to (4.38), where $\boldsymbol{Y}$ is given by (4.28), with the same form as in (4.36) and, corresponding to the critical value $\lambda = \lambda_{k,n}$ (where $\lambda_{k,n}$ is the $n$th critical point for out-of-plane null solution of $\mathfrak{L}(\lambda)$ for $\mathtt{D}_k$ symmetric modes; recall that $n = 1$ and it is suppressed in writing for $\lambda < 1$, and $n = 1, 2, \ldots$ for $\lambda > 1$), $\boldsymbol{X}_k$ is same as that stated as $\boldsymbol{X}_1$ in (A.1). Note, again, that $\boldsymbol{X}_k$ is normalized so that there is no arbitrary scalar in (4.38). The same expressions, as mentioned so far in this paragraph, hold for post-buckling analysis of $\mathtt{O}(2)$ symmetric solutions; this can be done by simply setting $k = 0$ and removing $\theta$ dependent terms.

The analysis presented in this subsection depends on the transversality condition being satisfied. The argument for the latter is not completely rigorous (recall Remarks 4.2, 4.3) but there exists no numerical evidence against it.

**5. Numerical results.** A symmetry-reduced finite element method [30] is used to validate the semi analytical local analysis. The projection operator theory (section 3.4 of [33]) is used to partition the finite element vector space in accordance with irreducible representations of $\mathtt{G}$. Some details are summarized in Appendix D. Hermite shape functions are used for radial interpolation, Fourier series is used for azimuthal interpolation and regularity conditions are applied as $r \to 0^+$. The reduction using the projections defined above allows for marching along the bifurcating branch bypassing the singularities. For numerical illustrations, we consider the following sets of structure parameters:

- The plate is made of copper and the rod is made of steel, with both having the same thickness. (A)
- The plate consists of a soft organic material (pumpkin), while the rod is composed of a stiffer pumpkin skin ([7]). Since the ratio of Young's modulus is provided in the reference, an arbitrary unit 'E' is used, which cancels out during the non-dimensionalization process. (B)
- A theoretical scenario where the Young's modulus of the plate is two order of magnitude lower than that of the rod. (C)
- A theoretical scenario where the Young's modulus of the plate is two order of magnitude higher than that of the rod. (D)

The choice of structure parameters for the four sets above is stated in Table 1 and Table 2.

The bending and torsional stiffness of the rod (for simplicity, the rod is assumed to have circular cross section), based on their values in engineering systems listed in

| Structure parameters | A | B | C | D |
|---|---|---|---|---|
| $E_{plate}$ | 110 GPa | 0.6 E | 1 GPa | 100 GPa |
| $\nu_{plate}$ | 0.34 | 0.5 | 0.3 | 0.3 |
| $E_{rod}$ | 200 GPa | 1 E | 100 GPa | 1 GPa |

Table 1: Physical structure parameter values considered for analysis. In all cases, $\nu_{rod} = 0.3$ is chosen as the Poisson's ration of the rod; $\mathfrak{D} = 0.028$m is the diameter of center-line of stress-free circular rod; $\mathtt{d}_r = h_{plate} = 0.002$m where $\mathtt{d}_r$ is the diameter of rod cross-section and $h_{plate}$ is the thickness of plate. Note that $\mathfrak{C}$ in Eq. (1.1) has the units Pa-m.

Table 1, are computed by applying the formulae [9]: $\beta = E_{rod}I_{rod}, \gamma = \frac{E_{rod}}{2(1+\nu_{rod})}J_{rod}$, where $I_{rod}$ and $J_{rod}$ represent the diametrical area moment of inertia and the polar moment of inertia for the circular cross-section, respectively. Using the non-dimensional scheme described in (1.1), and adopting the length scale $L = \mathfrak{D}/2$, the corresponding non-dimensional structure parameters are determined and are listed in Table 2. With this length scale, the parameter $\alpha$ is set to 1 for all configurations.

| Parameter values | A | B | C | D |
|---|---|---|---|---|
| $\beta$ | 0.00276149 | 0.00214668 | 0.156278 | 0.0000156278 |
| $\gamma$ | 0.00212422 | 0.00165129 | 0.120214 | 0.0000120214 |
| $\nu$ | 0.34 | 0.5 | 0.3 | 0.3 |

Table 2: Structure parameter sets A–D. In all cases $h = 0.142857$.

Henceforth, we discuss the plots corresponding to local bifurcation curves for the four sets of structure parameter values. Parametric sweep for the curvature coefficient of the bifurcation curve has been also carried out around each of the four structure parameter sets, the plots of which are provided in the supplementary S2. A shift in nature of bifurcation from supercritical to subcritical has been discovered from the parameter sweep study only. The deviation between bifurcation curve from local analysis and finite element scheme is illustrated for the parameter value corresponding to significant deviation. Shapes of solutions from both methods are also plotted.

Fig. 2 gives a summary view of critical points, null spaces and associated local bifurcation curves corresponding to parameter set A of Table 2. The symbol $\circ$ and the magenta curves represent bifurcation points and local bifurcation curves, respectively, corresponding to $\mathtt{D}_k$ symmetry for $\lambda < 1$, i.e., non-planar solutions when plate is in tension in its initial state. Only bifurcations corresponding to first four modes, i.e., $k = 2, 3, 4, 5$, are considered. The symbol $\times$ and the black curves are the bifurcation points and local bifurcation curves, respectively, corresponding to $\mathtt{Z}_k$ symmetry for $\lambda < 1$, i.e., planar solutions when plate is in tension. Only first eight modes, i.e., $k = 2, 3, \ldots, 8$, are considered. The symbol $\diamond$ and the red curves are the bifurcation points and local bifurcation curves, respectively, corresponding to $\mathtt{O}(2)$ symmetry and they lie on $\lambda > 1$ part of the plot. Only the bifurcation points with $\lambda < 1.2$ are considered. The symbol $*$ and the blue curves are the bifurcation points and the bifurcation curves, respectively, corresponding to $\mathtt{D}_k$ symmetry when $\lambda > 1$, i.e., plate is in compression in its initial state. Only bifurcations corresponding to $k = 2, 3, 4$ with bifurcation point $\lambda < 1.2$ are considered. Some of the buckling modes are shown and labelled with arrows to appropriate bifurcation points. The same colour scheme, as in Fig. 2, is followed for Fig. 3 which gives the bifurcation curves for all four structure parameter sets.

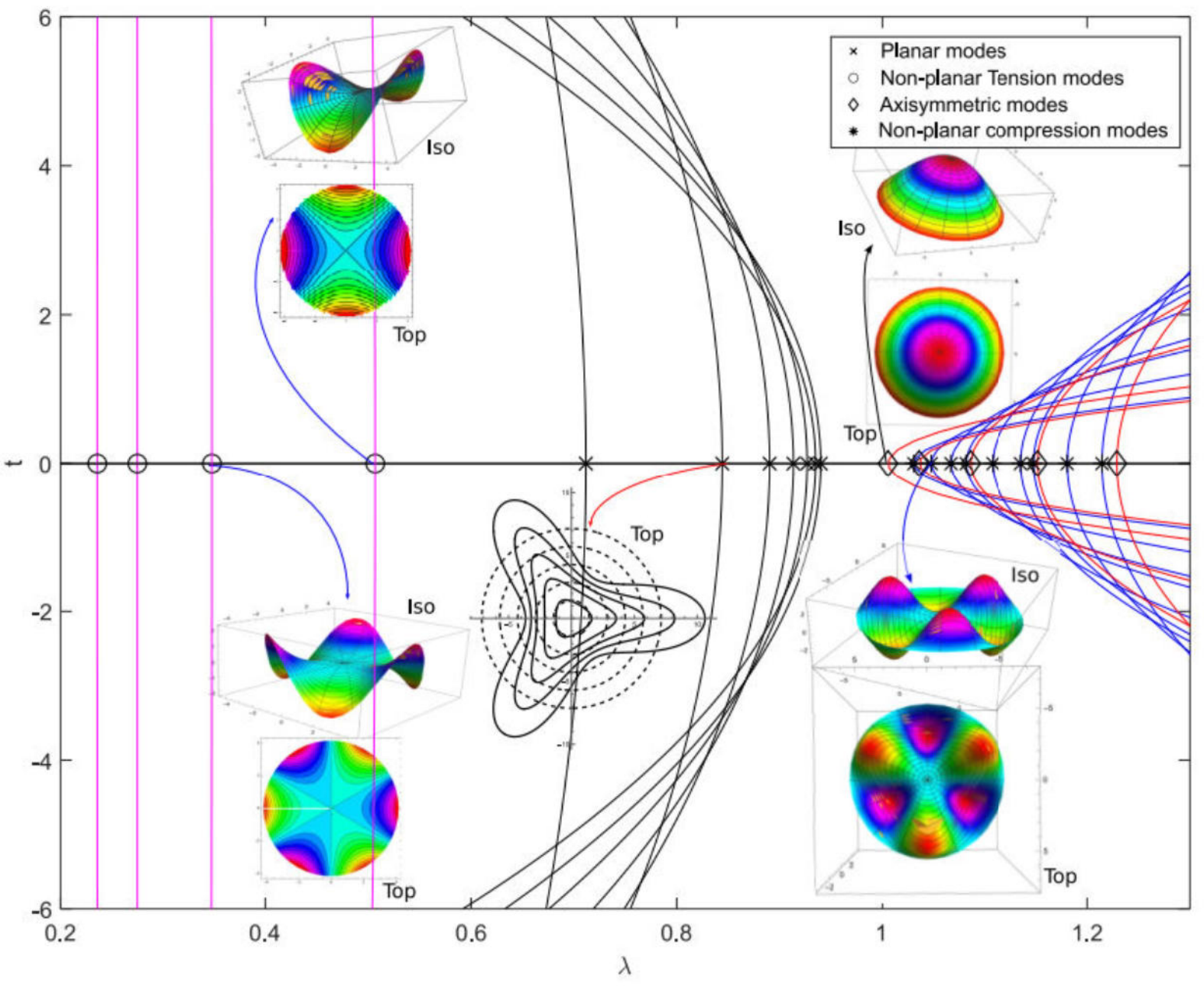

Fig. 2: Selected critical values and corresponding mode shapes for structure parameter set A. 'Iso' stands for isometric view, 'Top' stands for top view. Symbol $\circ$ and the magenta curves represent bifurcation points and local bifurcation curves, respectively, corresponding to $\mathtt{D}_k$ symmetry for $\lambda < 1$. Symbol $\times$ and the black curves are the bifurcation points and local bifurcation curves, respectively, corresponding to $\mathtt{Z}_k$ symmetry. Symbol $\diamond$ and the red curves are the bifurcation points and local bifurcation curves, respectively, corresponding to $\mathtt{O}(2)$ symmetry. Symbol $*$ and the blue curves are the bifurcation points and the bifurcation curves, respectively, corresponding to $\mathtt{D}_k$ symmetry when $\lambda > 1$.

Fig. 4a gives the shape of $\mathtt{Z}_3$ symmetric local nonlinear solution along with the null solution counterpart. The solid curve shows the post buckling solution shape and the dashed curve shows the null solution (mode) scaled by the bifurcation parameter. Both are plotted for bifurcation curve parameter $t = 3$ and the resulting deformation is scaled by a factor of 3. In Fig. 4b, the solid curve is the bifurcation curve obtained from local analysis and dashed curve is the bifurcation curve is obtained from the symmetry-reduced finite element method for $\mathtt{D}_k$ symmetry and $\lambda < 1$. It is observed that the bifurcation curve is indeed subcritical close to the critical point and deviates from the theoretical one farther away.

It has been found that all bifurcation curves in Fig. 5 are supercritical except for magenta curves corresponding to structure parameter set C. Also, it is found that the nature of bifurcation is not the same for all structure parameter values. To investigate this, a study involving the sweep of structure parameters is conducted for evaluation of $\ddot{\lambda}(0)$.[1] The colour scheme is same as the one used for bifurcation curves. The change of sign is observed in the structure parametric sweep corresponding to Poisson's ratio $\nu$ for structure parameter set C. A zoomed in view of it is also included in Fig. 5. A sign change in $\ddot{\lambda}(0)$ is observed only for magenta curves, i.e., $\mathtt{D}_k$ symmetric solution for $\lambda < 1$. It is observed that out of four bifurcation points under consideration, two are subcritical, i.e., $\ddot{\lambda}(0) > 0$ , on increasing the value of $\nu$, value of $\ddot{\lambda}(0)$ becomes closer to 0. At high enough value only one of four bifurcation points remain locally subcritical with small value of $\ddot{\lambda}(0)$. This indicates a crucial change in the nature of bifurcation. Symmetry based finite element analysis is carried out with structure parameter values of set C but with $\nu = 0.45$ to study the subcritical nature of bifurcation. This value has been selected as the value of $\ddot{\lambda}(0)$ corresponding to it is close to zero while remaining subcritical, as seen in Fig. 5 marked with a ×, therefore it is expected to deviate faster towards $\ddot{\lambda} < 0$ which is also a deviation from the theoretical results by local analysis.

[1]Some illustrative results for structure parameter sweeps are included in supplementary S2 and S3.

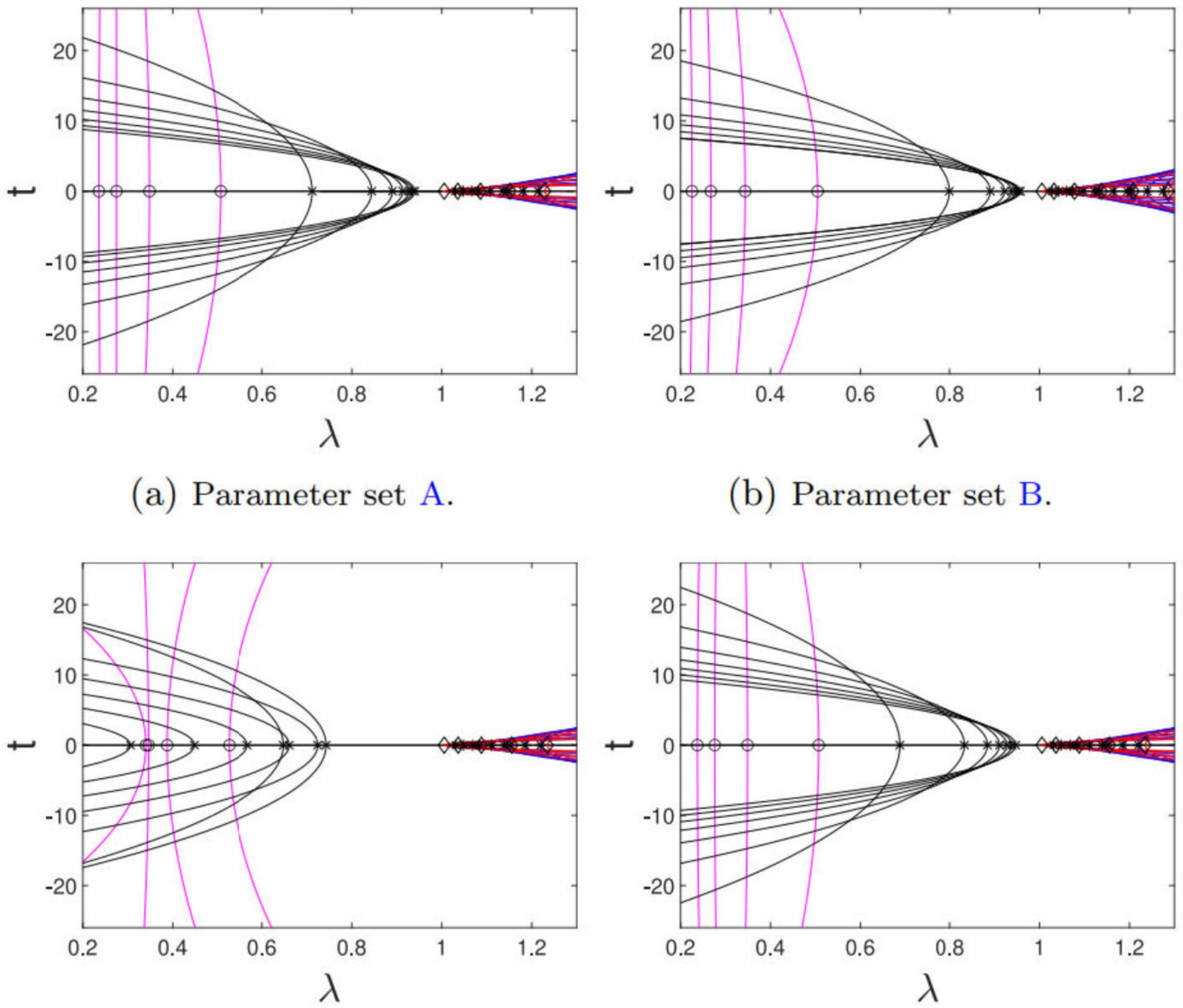

(a) Parameter set A. (b) Parameter set B.

(c) Parameter set C. (d) Parameter set D.

Fig. 3: Local Bifurcation curves for all four structure parameter sets. The symbols used are same as those described in caption of Fig. 2.

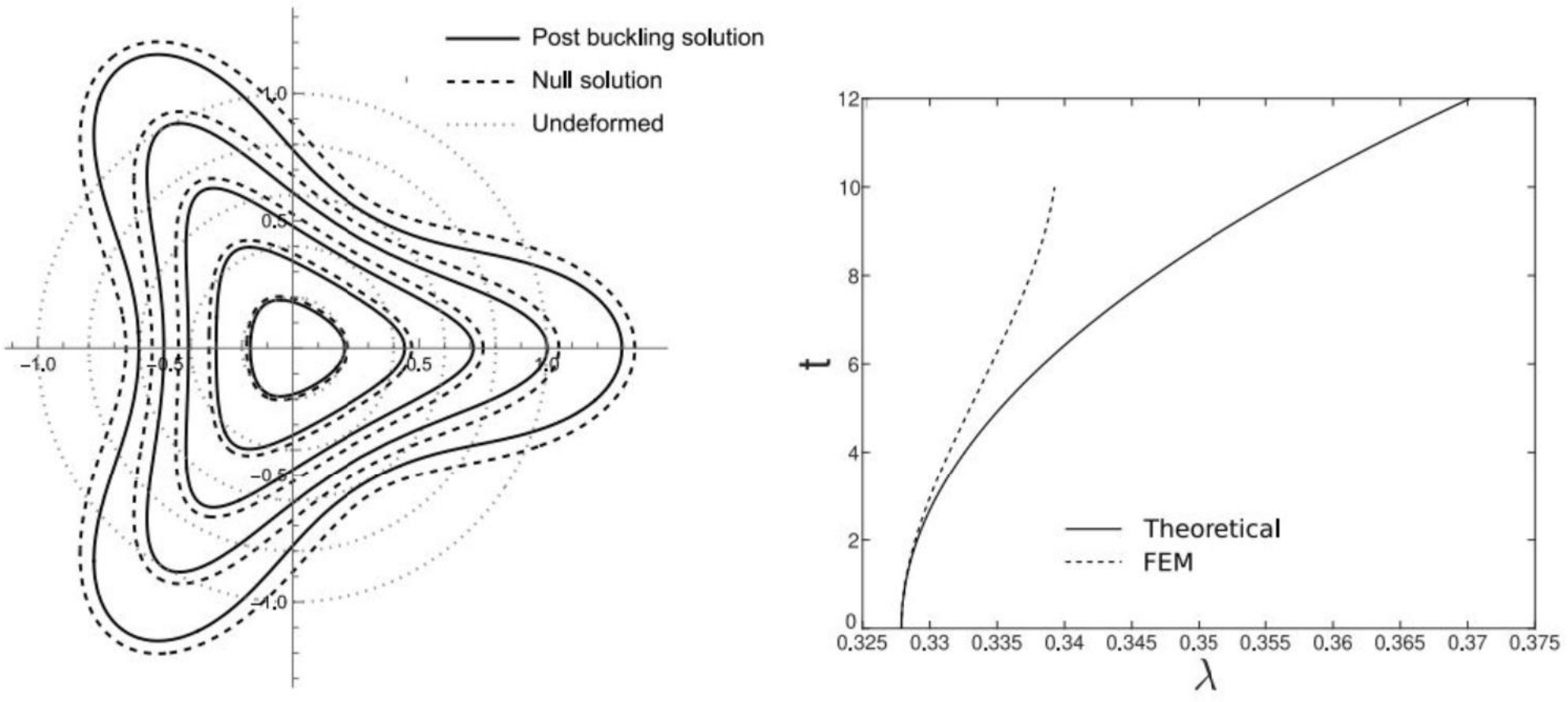

(a) (b)

Fig. 4: (a) Local post buckling solution for $\mathtt{Z}_3$ symmetry corresponding to structure parameter set A. ($t = 3$ and deformation scaled by a factor of 3). (b) Theoretical $(4.38)_2$ and finite element simulation based bifurcation curves at the peculiar point in structure parameter set C.

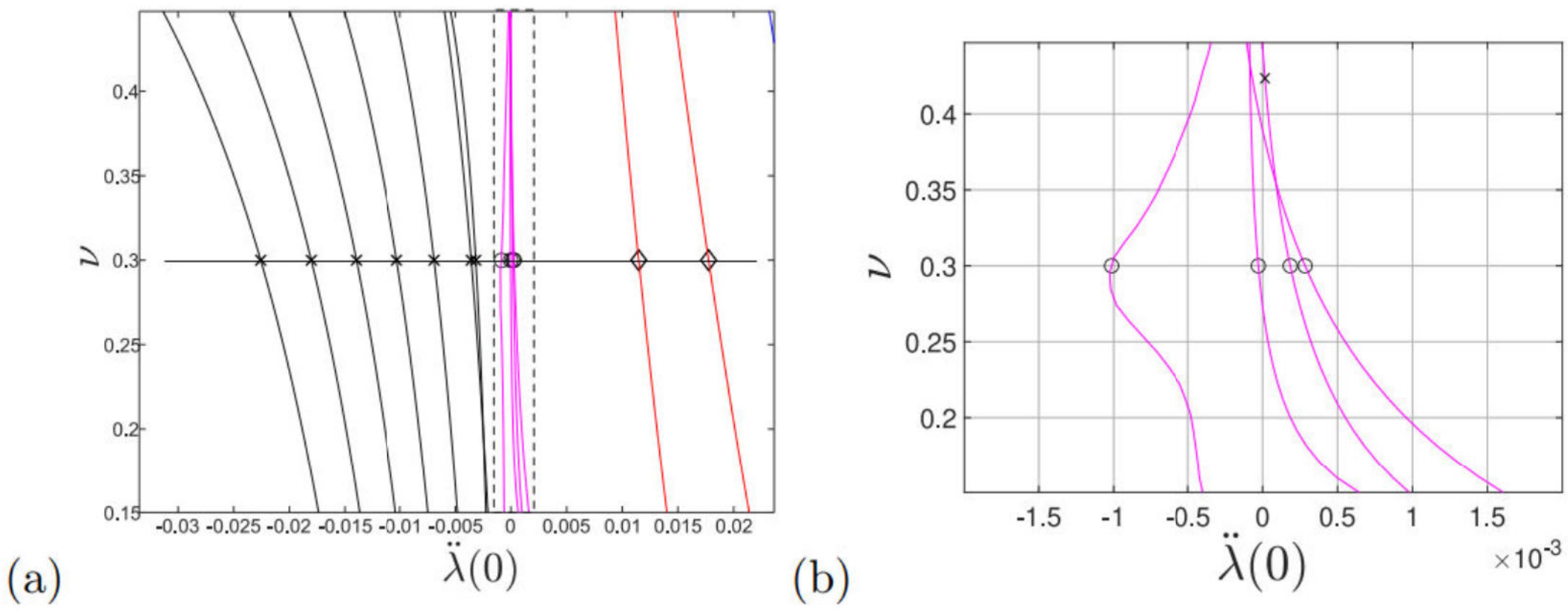

Fig. 5: (a) Parameter sweep around the structure parameter set C. Symbol ∘ and the magenta curves represent $\ddot{\lambda}(0)$ corresponding to $\lambda_c < 1$ of $\mathtt{D}_k$ symmetric branch. Symbol × and the black curves represent $\ddot{\lambda}(0)$ corresponding to $\lambda_c < 1$ of $\mathtt{Z}_k$ symmetric branch. Symbol ⋄ and the red curves represent $\ddot{\lambda}(0)$ corresponding to $\lambda_c > 1$ of $\mathtt{O}(2)$ symmetric branch. (b) The zoom-in section corresponding to the dashed rectangle in (a). Symbol × represent the point corresponding to the parameter value used for subsequent FEM analysis.

**6. Concluding remarks.** Based on the analysis, several key points can be concluded. As one of the main results, it is found that all critical points within the evaluated range are bifurcation points. In most cases, the bifurcation is supercritical for the given structure parameters, except when the plate material is much softer compared to the rod. When the plate becomes significantly softer, the bifurcation changes from supercritical to subcritical. It is also observed that the twisting stiffness of the rod has almost no effect on the shape of the bifurcation curves. This is plausible as the post-buckling behavior is mainly dominated by bending, with very little rod twisting involved. Furthermore, even though the symmetry-reduced finite element method is computationally very slow because it uses Fourier shape functions instead of polynomial one, it is still very helpful for finding solutions beyond the critical points in order to compare with theoretically obtained local bifurcation curves. The post-buckling analysis also allows us to perform parameter study which can be computationally expensive for direct numerical approach (using finite elements, say).

In this article, although we do not have any precise characterisation of the stability of bifurcating branches against any arbitrary perturbation, it is expected that supercritical bifurcations are associated with stable solution branches, while subcritical bifurcations are associated with unstable solution branches. Moreover, we can comment further on the local stability around $\lambda = 1$. For small perturbations the closest bifurcation points on either side of $\lambda = 1$ provide the bounds for which the trivial solution is stable. Based on our numerical evaluations of parameter sweeps of critical values of $\lambda$ (included as graphical plots in the supplementary S2), a lower bound of stability is provided by the largest $\mathtt{Z}_k$ symmetric bifurcation point and a upper bound is given by the lowest $\mathtt{O}(2)$ symmetric bifurcation point (bifurcation points with other symmetries are found to be farther away from $\lambda = 1$). We have the following expression for critical points $\lambda_k$ for $\mathtt{Z}_k$ symmetric solutions [5] ($\lambda < 1$):

$$\lambda_k = \frac{\begin{array}{c}48 + k(24 - 24\nu) + \beta\alpha^3 k^4(3 - \nu)\\ +k^2(-36 - 3\beta\alpha^3 - 24\nu + \beta\alpha^3\nu + 12\nu^2)\end{array}}{12(\nu + 1)(\nu - 3)(k^2 - 1)}, \quad k = 2, 3, \dots . \tag{6.1}$$

The planar modes exist only all those $k$ such that $\lambda_k \in (0, 1)$, $k = 2, 3, \dots$; as $k \to \infty$, the critical value $\lambda_k \to -\infty$, therefore the expression crosses 0 for some large but finite $k = k_{\max}$, therefore, there are a finite number of critical values $\lambda_k$. It can be easily shown that $\lambda_k$ (6.1) is strictly less than one for $k = 2, 3, \dots$, therefore there exists an upper bound on $\lambda_k$ which is less than 1. This gives the lower bound $\lambda_L$ in the interval of stability $(\lambda_L, \lambda_U)$ around $\lambda = 1$, namely,

$$\lambda_L = \max_{k=2,3,\dots,k_{\max}} \left( \frac{\begin{array}{c}48 + k(24 - 24\nu) + \beta\alpha^3 k^4(3 - \nu)\\ +k^2(-36 - 3\beta\alpha^3 - 24\nu + \beta\alpha^3\nu + 12\nu^2)\end{array}}{12(\nu + 1)(\nu - 3)(k^2 - 1)} \right). \tag{6.2}$$

Note that the $\mathtt{O}(2)$ symmetric solutions are identical to the buckled solutions for an equivalent simply supported plate as the rod only admits rigid body displacements in this symmetry. The $\mathtt{O}(2)$ symmetric (axisymmetric) solutions of simply supported circular plate are well studied [2]. Using the relation between $\lambda$ and the planar (homogeneous) stress, derived in [5], present in the plate for the trivial solution, and using the known value of the lowest plate buckling load from [2], the upper bound $\lambda_U$ ($> 1$), that is the lowest critical value for $\lambda > 1$, is found to be

$$\lambda_U = 1 + \frac{0.7h^2\alpha^2}{4(1 + \nu)}. \tag{6.3}$$

$\lambda_L$ (6.2) and $\lambda_U$ (6.3) are expected to serve as stability bounds around $\lambda = 1$ for small perturbations.

As usual in modeling physical problems, there are some limitations of the presented analysis. The von-Kármán plate model only considers small strain in membrane and linear out-of-plane dislacement-curvature relation, which makes it unsuitable for cases where plate's edge undergoes large three-dimensional deformation. Moreover, Kirchhoff assumption for the rod, although adds nonlinearity to the rod-plate model, limits the interaction between the rod and the plate, in the context of bifurcation modes for $\lambda < 1$, that is when the plate is homogeneously deformed (under constant tensile self-stress) in the trivial solution. Future research work may address these limitations by using shell models instead of plate models, which allows for a study of large deformation. The effect of twisting stiffness can also be enhanced by using chiral rod in place of achiral Kirchhoff rod, thus making the model more realistic.

**Acknowledgments.** The authors thank the anonymous referees for their constructive comments and suggestions that helped to improve the manuscript. In particular, the concluding remarks involving (6.2) and (6.3) have been included in response to a question raised by one of the reviewers. The authors thank Sanjay Dharmavaram for discussions during his visit to IIT Kanpur in December 2021.

**Appendix A. Null space of $\mathfrak{L}(\lambda)$.**

*Non-planar, non-axisymmetric.* In this case, the null solution has the form $\Big((0,0), y, [0,0,0], [M_{rr}, M_{\theta\theta}, M_{r\theta}], (f_r, f_\theta), (s_r, s_\theta, 0), (0,0,n_3)\Big)^\top$ with

$$\begin{aligned}
&\boldsymbol{X}(r,\theta) = A_k\Big((0,0), \phi(r)\cos k\theta, [0,0,0], [\Phi_{rr}(r)\cos k\theta, \Phi_{\theta\theta}(r)\cos k\theta, \Phi_{r\theta}(r)\sin k\theta],\\
&(\phi_{fr}(r)\cos k\theta, \phi_{f\theta}(r)\sin k\theta), (\phi_{sr}(r)\sin k\theta, \phi_{s\theta}(r)\cos k\theta, 0), (0,0,\phi_{n3}\sin k\theta)\Big)^\top\\
&+ B_k\Big((0,0), \psi(r)\sin k\theta, [0,0,0], [\Psi_{rr}(r)\sin k\theta, \Psi_{\theta\theta}(r)\sin k\theta, \Psi_{r\theta}(r)\cos k\theta],\\
&(\psi_{fr}(r)\sin k\theta, \psi_{f\theta}(r)\cos k\theta), (\psi_{sr}(r)\cos k\theta, \psi_{s\theta}(r)\sin k\theta, 0), (0,0,\psi_{n3}\cos k\theta)\Big)^\top\\
&= \boldsymbol{X}_1(r,\theta) + \boldsymbol{X}_2(r,\theta),
\end{aligned} \tag{A.1}$$

where $\phi, \psi, \Phi_{rr}, \ldots$ are explicitly found; again, due their cumbersome expressions we do not include them in this main text.

*Planar.* In this case, the null solution has the form $\Big((v_r, v_\theta), 0, [N_{rr}, N_{\theta\theta}, N_{r\theta}], [0,0,0], (0,0), (0,0,s_3), (n_r, n_\theta, 0)\Big)^\top$ with

$$\begin{aligned}
&\boldsymbol{X}(r,\theta)\\
&= A_k\Big((\phi_r(r)\cos k\theta, \phi_\theta(r)\sin k\theta), 0, [\Phi_{rr}(r)\cos k\theta, \Phi_{\theta\theta}(r)\cos k\theta, \Phi_{r\theta}(r)\sin k\theta],\\
&[0,0,0], (0,0), (0,0,\phi_{s3}\sin k\theta), (\phi_{nr}\sin k\theta, \phi_{n\theta}\cos k\theta, 0)\Big)^\top +\\
&+ B_k\Big((\psi_r(r)\sin k\theta, \psi_\theta(r)\cos k\theta), 0, [\Psi_{rr}(r)\sin k\theta, \Psi_{\theta\theta}(r)\sin k\theta, \Psi_{r\theta}(r)\cos k\theta],\\
&[0,0,0], (0,0), (0,0,\psi_{s3}\cos k\theta), (\psi_{nr}\cos k\theta, \psi_{n\theta}\sin k\theta, 0)\Big)^\top\\
&= \boldsymbol{X}_1(r,\theta) + \boldsymbol{X}_2(r,\theta),
\end{aligned} \tag{A.2}$$

where $\phi_r, \psi_r, \Phi_{rr}, \ldots$ are explicitly found but due their cumbersome expressions we do not include them in the main text in this article.

*Axisymmetric.* In this case, the null solution has the form $\Big((0,0), y, [0,0,0], [M_{rr}, M_{\theta\theta}, 0], (f_r, 0), (0,0,0), (0,0,0)\Big)^\top$ with

$$\boldsymbol{X}(r,\theta) = C\Big((0,0), \phi(r), [0,0,0], [\Phi_{rr}(r), \Phi_{\theta\theta}(r), 0], (\phi_{f_r}(r), 0), (0,0,0), (0,0,0)\Big)^\top, \tag{A.3}$$

where $\phi, \Phi_{rr}, \ldots$ are explicitly found but due their cumbersome expressions we do not state in this article.

**Appendix B. Adjoint operator $\mathfrak{L}^*$.**

We consider the definition of $\boldsymbol{Z}$ in (1.7), which is the direct notation for $\boldsymbol{Z} = \Big((u_r, u_\theta), w, [S_{rr}, S_{\theta\theta}, S_{r\theta}], [T_{rr}, T_{\theta\theta}, T_{r\theta}], (p_r, p_\theta), (q_r, q_\theta, q_3), (h_r, h_\theta, h_3)\Big)^\top \in \mathcal{R}$ (analogous to the definitions in §3.2). The adjoint of operator $\mathfrak{L}$, denoted by $\mathfrak{L}^*$, is defined by the relation $\langle \mathfrak{L}(\boldsymbol{X}), \boldsymbol{Z}\rangle = \langle \boldsymbol{X}, \mathfrak{L}^*(\boldsymbol{Z})\rangle$, for all $\boldsymbol{X} \in \mathcal{D}$ and $\boldsymbol{Z} \in \mathcal{R} \subset \mathcal{C}$, where $\langle\cdot,\cdot\rangle$ is the $L^2$ inner-product

$$\begin{aligned}
\langle \boldsymbol{X}, \boldsymbol{Y}\rangle = &\textstyle\int_\Omega (\boldsymbol{x}_1\cdot\boldsymbol{y}_1 + x_2 y_2 + \mathbf{X}_3 : \mathbf{Y}_3 + \mathbf{X}_4 : \mathbf{Y}_4 + \boldsymbol{y}_5\cdot\boldsymbol{y}_5)\ \mathrm{d}A +\\
&+ \textstyle\int_\Gamma (\boldsymbol{x}_6\cdot\boldsymbol{y}_6 + \boldsymbol{x}_7\cdot\boldsymbol{y}_7)\,\mathrm{d}l,
\end{aligned} \tag{B.1}$$

with $\boldsymbol{X} = \Big(\boldsymbol{x}_1, x_2, \mathbf{X}_3, \mathbf{X}_4, \boldsymbol{x}_5, \boldsymbol{x}_6, \boldsymbol{x}_7\Big)^\top$, $\boldsymbol{Y} = \Big(\boldsymbol{y}_1, y_2, \mathbf{Y}_3, \mathbf{Y}_4, \boldsymbol{y}_5, \boldsymbol{y}_6, \boldsymbol{y}_7\Big)^\top$. Hence, when $\boldsymbol{Z}$ is a null solutions of $\mathfrak{L}^*$, we get $\langle \mathfrak{L}(\boldsymbol{X}), \boldsymbol{Z}\rangle = 0, \forall \boldsymbol{X} \in \mathcal{U}$. Substituting (2.1) in left side of $\langle \mathfrak{L}(\boldsymbol{X}), \boldsymbol{Z}\rangle = \langle \boldsymbol{X}, \mathfrak{L}^*(\boldsymbol{Z})\rangle$, applying integration by parts appropriately, we get to the right side, for arbitrary $\boldsymbol{X}$, as:

$$
\begin{aligned}
&\langle \mathcal{L}(\boldsymbol{X}), \boldsymbol{Z}\rangle \\
&= \textstyle\int_\Omega \boldsymbol{v} \cdot \Big( -12(1-\nu)\nabla\cdot\mathbf{S} - 12\nu\nabla(\mathbf{S}:\mathbf{I})\Big)\mathrm{d}A \\
&\quad + \textstyle\int_\Omega z\Big( -\frac{h^2}{\lambda^2}(1-\nu)\nabla\cdot(\nabla\cdot\mathbf{T}) - \frac{h^2}{\lambda^2}\nu\Delta(\mathbf{T}:\mathbf{I}) + \frac{12}{\lambda}(1+\nu)(1-\lambda)\nabla\cdot\boldsymbol{p}\Big)\mathrm{d}A \\
&\quad + \textstyle\int_\Omega \mathbf{N} : (-\nabla\boldsymbol{u} - \mathbf{S})\mathrm{d}A + \int_\Omega \mathbf{M} : (\nabla^2 w - \mathbf{T})\mathrm{d}A + \int_\Omega \boldsymbol{f}\cdot(-\nabla w + \boldsymbol{p})\mathrm{d}A \\
&\quad + \textstyle\int_\Gamma \mathbf{s}\cdot\Big(2\boldsymbol{q}'' - 2(\gamma-1)\big(\frac{8}{\mathfrak{D}^2}(\boldsymbol{t}\otimes\boldsymbol{t} - \boldsymbol{l}\otimes\boldsymbol{l})\cdot\boldsymbol{q} \\
&\quad +2\alpha(\boldsymbol{l}\otimes\boldsymbol{t} + \boldsymbol{t}\otimes\boldsymbol{l})\cdot\boldsymbol{q}' - (\boldsymbol{t}\otimes\boldsymbol{t})\cdot\boldsymbol{q}''\big) \\
&\quad + 2\alpha(\gamma-\beta)\left(2\alpha(\boldsymbol{t}\otimes\boldsymbol{t} - \boldsymbol{l}\otimes\boldsymbol{l})\boldsymbol{q} + (\boldsymbol{t}\otimes\boldsymbol{l} + \boldsymbol{l}\otimes\boldsymbol{t})\boldsymbol{q}'\right) - 2\boldsymbol{t}\wedge\boldsymbol{h}\Big)\mathrm{d}l \qquad \text{(B.2)} \\
&\quad + \textstyle\int_\Gamma \boldsymbol{n}\cdot(-\boldsymbol{u}' - w'\boldsymbol{e}_3 + \boldsymbol{q}\wedge\boldsymbol{t})\mathrm{d}l + \int_\Gamma \boldsymbol{v}\cdot\Big(12(1-\nu)\mathbf{S}\boldsymbol{l} + 12\nu(\mathbf{S}:\mathbf{I})\boldsymbol{l} \\
&\quad + \frac{12}{\alpha}(1+\nu)(\lambda-1)(\boldsymbol{q}'\wedge\boldsymbol{t} - \alpha\boldsymbol{q}\wedge\boldsymbol{l}) - \boldsymbol{h}'\Big)\mathrm{d}l \\
&\quad + \textstyle\int_\Gamma z\Big(\frac{h^2}{\lambda^2}\nu\nabla(\mathbf{T}:\mathbf{I})\cdot\boldsymbol{l} + \frac{h^2}{\lambda^2}(1-\nu)(\nabla(T_{r\theta})\cdot\boldsymbol{t} + (\nabla\cdot\mathbf{T})\cdot\boldsymbol{l}) \\
&\quad - \frac{12}{\lambda}(1+\nu)(1-\lambda)\boldsymbol{p}\cdot\boldsymbol{l} + \frac{12}{\alpha}(1+\nu)(\lambda-1)(\alpha\boldsymbol{t}\cdot\boldsymbol{q} + \boldsymbol{l}\cdot\boldsymbol{q}') - \boldsymbol{e}_3\cdot\boldsymbol{h}'\Big)\mathrm{d}l \\
&\quad - \frac{h^2}{\lambda^2}\textstyle\int_\Gamma \nabla z\cdot\boldsymbol{l}((1-\nu)T_{rr} + \nu\mathbf{T}:\mathbf{I})\mathrm{d}l.
\end{aligned}
$$

Eventually, we obtain the components of $\mathfrak{L}^*(\lambda)$ and we get the following linear boundary restrictions on $\boldsymbol{Z}$ (1.7), useful to define the dual space:

$$
\begin{aligned}
12(1-\nu)\mathbf{S}\boldsymbol{l} + \frac{12}{\alpha}(1+\nu)(\lambda-1)((\mathtt{I} - \boldsymbol{e}_3\otimes\boldsymbol{e}_3)(\boldsymbol{q}'\wedge\boldsymbol{t}) + \\
+12\nu(\mathbf{S}:\mathbf{I})\boldsymbol{l} - \alpha(\mathtt{I} - \boldsymbol{e}_3\otimes\boldsymbol{e}_3)(\boldsymbol{q}\wedge\boldsymbol{l})) - (\mathtt{I} - \boldsymbol{e}_3\otimes\boldsymbol{e}_3)\boldsymbol{h}' = \mathbf{0}, \quad \text{(B.3a)}
\end{aligned}
$$

$$
\begin{aligned}
\frac{h^2}{\lambda^2}(1-\nu)(\nabla(\boldsymbol{e}_\theta\cdot\mathbf{T}\boldsymbol{e}_r)\cdot\boldsymbol{t} + (\nabla\cdot\mathbf{T})\cdot\boldsymbol{l}) - \frac{12}{\lambda}(1+\nu)(1-\lambda)\boldsymbol{p}\cdot\boldsymbol{l} - \\
-\boldsymbol{e}_3\cdot\boldsymbol{h}' + \frac{h^2\nu}{\lambda^2}\nabla(\mathbf{T}:\mathbf{I})\cdot\boldsymbol{l} + \frac{12}{\alpha}(1+\nu)(\lambda-1)(\alpha\boldsymbol{t}\cdot\boldsymbol{q} + \boldsymbol{l}\cdot\boldsymbol{q}') = 0, \quad \text{(B.3b)}
\end{aligned}
$$

$$
\boldsymbol{e}_r\cdot\mathbf{T}\boldsymbol{e}_r + \nu\boldsymbol{e}_\theta\cdot\mathbf{T}\boldsymbol{e}_\theta = 0. \quad \text{(B.3c)}
$$

The adjoint $\mathfrak{L}^*(\lambda)$ of the linear operator $\mathfrak{L}(\lambda)$ is given by

$$
\mathfrak{L}^*(\lambda)[\boldsymbol{Z}] = \begin{pmatrix}
-12(1-\nu)\nabla\cdot\mathbf{S} - 12\nu\nabla(\mathbf{S}:\mathbf{I}) \\
-\frac{h^2}{\lambda^2}(1-\nu)\nabla\cdot(\nabla\cdot\mathbf{T}) - \frac{h^2}{\lambda^2}\nu\Delta(\mathbf{T}:\mathbf{I}) + \frac{12}{\lambda}(1+\nu)(1-\lambda)\nabla\cdot\boldsymbol{p} \\
-\frac{1}{2}(\nabla\boldsymbol{u} + \nabla\boldsymbol{u}^\top) - \mathbf{S} \\
\nabla^2 w - \mathbf{T} \\
\boldsymbol{p} - \nabla w \\
2\mathbf{C}\boldsymbol{q}'' - 2\alpha(\gamma-1)(\boldsymbol{t}\otimes\boldsymbol{l} + \boldsymbol{l}\otimes\boldsymbol{t})\boldsymbol{q}' - 2\boldsymbol{t}\wedge\boldsymbol{h} \\
\boldsymbol{q}\wedge\boldsymbol{t} - \boldsymbol{u}' - w'\boldsymbol{e}_3
\end{pmatrix}, \quad \text{(B.4)}
$$

where $\mathbf{C}$ is defined by $(2.1)_2$, and $\boldsymbol{Z}$ is defined by (1.7) subject to the restrictions (B.3).

**Appendix C. Null solution of $\mathfrak{L}^*_{\mathsf{Z}_k}$.** The components of $\boldsymbol{Z}_k$ are

$$u_r(r,\theta) = (\frac{\zeta^2(k-1)r^2((k+2)\nu+k-2)}{(k+1)((k-2)\nu+k+2)} - 1)Ar^{k-1}\cos(k\theta), \tag{C.1a}$$

$$u_\theta(r,\theta) = (1 - \frac{\zeta^2(k-1)r^2(k\nu+k+4)}{(k+1)((k-2)\nu+k+2)})Ar^{k-1}\sin(k\theta), \tag{C.1b}$$

$$S_{rr}(r,\theta) = -\frac{(k-1)(k(\nu+1)(\alpha^2r^2-1)+2(\nu-1)(\alpha^2r^2+1))}{(k-2)\nu+k+2}Ar^{k-2}\cos(k\theta), \tag{C.1c}$$

$$S_{r\theta}(r,\theta) = \frac{(k-1)(k(\nu+1)(\alpha^2r^2-1)+2(\nu-1))}{(k-2)\nu+k+2}Ar^{k-2}\sin(k\theta), \tag{C.1d}$$

$$S_{\theta\theta}(r,\theta) = (k-1)r^{k-2}(\alpha^2r^2-1)Ar^{k-2}\cos(kt), \tag{C.1e}$$

$$q_3(\theta) = \frac{2(k-1)(\nu-3)(\frac{1}{\alpha})^{k-2}}{(k-2)\nu+k+2}A\sin(k\theta), \tag{C.1f}$$

$$h_r(\theta) = \frac{2(k-1)k^2(\nu-3)(\frac{1}{\alpha})^{k-4}}{(k-2)\nu+k+2}A\sin(k\theta), \tag{C.1g}$$

$$h_\theta(\theta) = \frac{24(\nu-1)(\frac{1}{\alpha})^{k-1}(k(\nu-1)-2)}{(k+1)((k-2)\nu+k+2)}A\cos(k\theta). \tag{C.1h}$$

**Appendix D. Symmetry-reduced finite element formulation.** A numerical finite element scheme is discussed based on [30]. Relevant parts from the cited work are summarised describing the features of the numerical scheme and modifications, specific to the problem of this aritcle. The structure parameters of the numerical scheme are stated. The following irreducible representation of complete symmetry group $\mathsf{G}$ is used to construct the required projections:

$$\mathrm{g}_1^{(1)} = \mathrm{g}_2^{(1)} = \mathrm{g}_3^{(1)} = \mathrm{g}_4^{(1)} = 1, \qquad \mathrm{g}_1^{(2)} = \mathrm{g}_2^{(2)} = -\mathrm{g}_3^{(2)} = -\mathrm{g}_4^{(2)} = 1, \tag{D.1}$$

$$\mathrm{g}_1^{(3)} = -\mathrm{g}_2^{(3)} = \mathrm{g}_3^{(3)} = -\mathrm{g}_4^{(3)} = 1, \qquad \mathrm{g}_1^{(4)} = -\mathrm{g}_2^{(4)} = -\mathrm{g}_3^{(4)} = \mathrm{g}_4^{(4)} = 1, \tag{D.2}$$

$$\mathrm{g}_1^{(2h+3)} = \begin{bmatrix} c_h & s_h \\ -s_h & c_h \end{bmatrix}, \ \mathrm{g}_2^{(2h+3)} = \begin{bmatrix} c_h & s_h \\ s_h & -c_h \end{bmatrix}, \mathrm{g}_3^{(2h+3)} = \begin{bmatrix} c_h & s_h \\ s_h & -c_h \end{bmatrix}, \tag{D.3}$$

$$\mathrm{g}_4^{(2h+3)} = \begin{bmatrix} c_h & s_h \\ -s_h & c_h \end{bmatrix}, \mathrm{g}_1^{(2h+4)} = \begin{bmatrix} c_h & s_h \\ -s_h & c_h \end{bmatrix}, \mathrm{g}_2^{(2h+4)} = \begin{bmatrix} c_h & s_h \\ s_h & -c_h \end{bmatrix}, \tag{D.4}$$

$$\mathrm{g}_3^{(2h+4)} = \begin{bmatrix} -c_h & -s_h \\ -s_h & c_h \end{bmatrix}, \ \mathrm{g}_4^{(2h+3)} = \begin{bmatrix} -c_h & -s_h \\ s_h & -c_h \end{bmatrix}, \tag{D.5}$$

using the shorthand $c_h = \cos h\theta, s_h = \sin h\theta$, and the corresponding projections for partitioning the space are

$$\mathcal{P}^1 = \frac{1}{8\pi}\int_{-\pi}^{\pi}[\mathcal{T}_{r_\theta} + \mathcal{T}_{er_\theta} + \mathcal{T}_{fr_\theta} + \mathcal{T}_{fer_\theta}]\mathrm{d}\theta, \tag{D.6}$$

$$\mathcal{P}^2 = \frac{1}{8\pi}\int_{-\pi}^{\pi}[\mathcal{T}_{r_\theta} + \mathcal{T}_{er_\theta} - \mathcal{T}_{fr_\theta} - \mathcal{T}_{fer_\theta}]\mathrm{d}\theta, \tag{D.7}$$

$$\mathcal{P}^3 = \frac{1}{8\pi}\int_{-\pi}^{\pi}[\mathcal{T}_{r_\theta} - \mathcal{T}_{er_\theta} + \mathcal{T}_{fr_\theta} - \mathcal{T}_{fer_\theta}]\mathrm{d}\theta, \tag{D.8}$$

$$\mathcal{P}^4 = \frac{1}{8\pi}\int_{-\pi}^{\pi}[\mathcal{T}_{r_\theta} - \mathcal{T}_{er_\theta} - \mathcal{T}_{fr_\theta} + \mathcal{T}_{fer_\theta}]\mathrm{d}\theta, \tag{D.9}$$

$$\mathcal{P}_1^{2h+3} = \frac{2}{8\pi}\int_{-\pi}^{\pi} c_h(\mathcal{T}_{r_\theta} + \mathcal{T}_{er_\theta} + \mathcal{T}_{fr_\theta} + \mathcal{T}_{fer_\theta})\mathrm{d}\theta, \tag{D.10}$$

$$\mathcal{P}_2^{2h+3} = \frac{2}{8\pi}\int_{-\pi}^{\pi} s_h(-\mathcal{T}_{r_\theta} + \mathcal{T}_{er_\theta} + \mathcal{T}_{fr_\theta} - \mathcal{T}_{fer_\theta})\mathrm{d}\theta, \tag{D.11}$$

$$\mathcal{P}_1^{2h+4} = \frac{2}{8\pi}\int_{-\pi}^{\pi} c_h(\mathcal{T}_{r_\theta} + \mathcal{T}_{er_\theta} - \mathcal{T}_{fr_\theta} - \mathcal{T}_{fer_\theta})\mathrm{d}\theta, \tag{D.12}$$

$$\mathcal{P}_2^{2h+4} = \frac{2}{8\pi}\int_{-\pi}^{\pi} s_h(-\mathcal{T}_{r_\theta} + \mathcal{T}_{er_\theta} - \mathcal{T}_{fr_\theta} + \mathcal{T}_{fer_\theta})\mathrm{d}\theta, \tag{D.13}$$

where $h \in \mathbb{N}$ and $\mathcal{T}_g$ is the symmetry operation on discretized variables, corresponding to group element $g \in \mathtt{G}$.